\documentclass[aps,prb,showpacs,superscriptaddress,numerical,reprint]{revtex4-1}
\usepackage{graphicx}
\usepackage{amsmath}
\usepackage{amssymb}
\usepackage{siunitx}
\usepackage{color}
\usepackage[colorlinks=true,citecolor=blue,linkcolor=blue,pdfstartview=FitH,pdfauthor=Gross]{hyperref}

\begin{document}
\title[Tunable coupling of transmission-line microwave resonators mediated by an rf~SQUID]
      {Tunable coupling of transmission-line microwave resonators mediated by an rf~SQUID}

\author{F. Wulschner}\email{Friedrich.Wulschner@wmi.badw-muenchen.de}

\affiliation{Walther-Mei{\ss}ner-Institut, Bayerische Akademie der Wissenschaften, D-85748 Garching, Germany}
\affiliation{Physik-Department, Technische Universit\"{a}t M\"{u}nchen, D-85748 Garching, Germany}

\author{J.~Goetz}
\affiliation{Walther-Mei{\ss}ner-Institut, Bayerische Akademie der Wissenschaften, D-85748 Garching, Germany}
\affiliation{Physik-Department, Technische Universit\"{a}t M\"{u}nchen, D-85748 Garching, Germany}

\author{F.~R.~Koessel}
\affiliation{Walther-Mei{\ss}ner-Institut, Bayerische Akademie der Wissenschaften, D-85748 Garching, Germany}
\affiliation{Physik-Department, Technische Universit\"{a}t M\"{u}nchen, D-85748 Garching, Germany}

\author{E.~Hoffmann}
\affiliation{Walther-Mei{\ss}ner-Institut, Bayerische Akademie der Wissenschaften, D-85748 Garching, Germany}
\affiliation{Physik-Department, Technische Universit\"{a}t M\"{u}nchen, D-85748 Garching, Germany}

\author{A.~Baust}
\affiliation{Walther-Mei{\ss}ner-Institut, Bayerische Akademie der Wissenschaften, D-85748 Garching, Germany}
\affiliation{Physik-Department, Technische Universit\"{a}t M\"{u}nchen, D-85748 Garching, Germany}
\affiliation{Nanosystems Initiative Munich (NIM), Schellingstra{\ss}e 4, 80799 M\"{u}nchen, Germany}

\author{P.~Eder}
\affiliation{Walther-Mei{\ss}ner-Institut, Bayerische Akademie der Wissenschaften, D-85748 Garching, Germany}
\affiliation{Physik-Department, Technische Universit\"{a}t M\"{u}nchen, D-85748 Garching, Germany}
\affiliation{Nanosystems Initiative Munich (NIM), Schellingstra{\ss}e 4, 80799 M\"{u}nchen, Germany}

\author{M.~Fischer}
\affiliation{Walther-Mei{\ss}ner-Institut, Bayerische Akademie der Wissenschaften, D-85748 Garching, Germany}
\affiliation{Physik-Department, Technische Universit\"{a}t M\"{u}nchen, D-85748 Garching, Germany}

\author{M.~Haeberlein}
\affiliation{Walther-Mei{\ss}ner-Institut, Bayerische Akademie der Wissenschaften, D-85748 Garching, Germany}
\affiliation{Physik-Department, Technische Universit\"{a}t M\"{u}nchen, D-85748 Garching, Germany}

\author{M.~J.~Schwarz}
\affiliation{Walther-Mei{\ss}ner-Institut, Bayerische Akademie der Wissenschaften, D-85748 Garching, Germany}
\affiliation{Physik-Department, Technische Universit\"{a}t M\"{u}nchen, D-85748 Garching, Germany}
\affiliation{Nanosystems Initiative Munich (NIM), Schellingstra{\ss}e 4, 80799 M\"{u}nchen, Germany}

\author{M.~Pernpeintner}
\affiliation{Walther-Mei{\ss}ner-Institut, Bayerische Akademie der Wissenschaften, D-85748 Garching, Germany}
\affiliation{Physik-Department, Technische Universit\"{a}t M\"{u}nchen, D-85748 Garching, Germany}
\affiliation{Nanosystems Initiative Munich (NIM), Schellingstra{\ss}e 4, 80799 M\"{u}nchen, Germany}

\author{E.~Xie}
\affiliation{Walther-Mei{\ss}ner-Institut, Bayerische Akademie der Wissenschaften, D-85748 Garching, Germany}
\affiliation{Physik-Department, Technische Universit\"{a}t M\"{u}nchen, D-85748 Garching, Germany}

\author{L.~Zhong}
\affiliation{Walther-Mei{\ss}ner-Institut, Bayerische Akademie der Wissenschaften, D-85748 Garching, Germany}
\affiliation{Physik-Department, Technische Universit\"{a}t M\"{u}nchen, D-85748 Garching, Germany}
\affiliation{Nanosystems Initiative Munich (NIM), Schellingstra{\ss}e 4, 80799 M\"{u}nchen, Germany}

\author{C.~W.~Zollitsch}
\affiliation{Walther-Mei{\ss}ner-Institut, Bayerische Akademie der Wissenschaften, D-85748 Garching, Germany}
\affiliation{Physik-Department, Technische Universit\"{a}t M\"{u}nchen, D-85748 Garching, Germany}

\author{B.~Peropadre}
\affiliation{Department of Chemistry and Chemical Biology, Harvard University, Cambridge, Massachusetts 02138, United States}

\author{J.-J.~Garcia Ripoll}
\affiliation{Instituto de Fisica Fundamental, IFF-CSIC, Calle Serrano 113b, Madrid E-28006, Spain}

\author{E.~Solano}
\affiliation{Physik-Department, Technische Universit\"{a}t M\"{u}nchen, D-85748 Garching, Germany}
\affiliation{Department of Physical Chemistry, University of the Basque Country UPV/EHU, Apartado 644, E-48080 Bilbao, Spain}
\affiliation{IKERBASQUE, Basque Foundation for Science, Maria Diaz de Haro 3, E-48013 Bilbao, Spain}

\author{K.~Fedorov}
\affiliation{Walther-Mei{\ss}ner-Institut, Bayerische Akademie der Wissenschaften, D-85748 Garching, Germany}
\affiliation{Physik-Department, Technische Universit\"{a}t M\"{u}nchen, D-85748 Garching, Germany}

\author{E.~P.~Menzel}
\affiliation{Walther-Mei{\ss}ner-Institut, Bayerische Akademie der Wissenschaften, D-85748 Garching, Germany}
\affiliation{Physik-Department, Technische Universit\"{a}t M\"{u}nchen, D-85748 Garching, Germany}

\author{F.~Deppe}
\affiliation{Walther-Mei{\ss}ner-Institut, Bayerische Akademie der Wissenschaften, D-85748 Garching, Germany}
\affiliation{Physik-Department, Technische Universit\"{a}t M\"{u}nchen, D-85748 Garching, Germany}
\affiliation{Nanosystems Initiative Munich (NIM), Schellingstra{\ss}e 4, 80799 M\"{u}nchen, Germany}

\author{A.~Marx}
\affiliation{Walther-Mei{\ss}ner-Institut, Bayerische Akademie der Wissenschaften, D-85748 Garching, Germany}

\author{R.~Gross}\email{Rudolf.Gross@wmi.badw-muenchen.de}
\affiliation{Walther-Mei{\ss}ner-Institut, Bayerische Akademie der Wissenschaften, D-85748 Garching, Germany}
\affiliation{Physik-Department, Technische Universit\"{a}t M\"{u}nchen, D-85748 Garching, Germany}
\affiliation{Nanosystems Initiative Munich (NIM), Schellingstra{\ss}e 4, 80799 M\"{u}nchen, Germany}

\date{\today}

\begin{abstract}

We realize tunable coupling between two superconducting transmission line resonators. The coupling is mediated by a non-hysteretic rf~SQUID acting as a flux-tunable mutual inductance between the resonators. From the mode distance observed in spectroscopy experiments, we derive a coupling strength $g/2\pi$ ranging between $\SI{-320}{\mega \hertz}$ and $\SI{37}{\mega \hertz}$. In the case of $g\,{\approx}\,0$ the microwave power  cross transmission between the two resonators can be reduced by almost four orders of magnitude compared to the case where the coupling is switched on. In addition, we observe parametric amplification by applying a suitable additional drive tone.

\end{abstract}

%\pacs{pacs}

\maketitle

\section{Introduction}

In circuit quantum electrodynamics, the controllable interaction of circuit elements is a highly desirable resource for quantum computation and quantum simulation experiments. The most common method is a static capacitive or inductive coupling between cavities and/or qubits. In such a system, photon exchange can be controlled by either tuning the circuit elements in and out of resonance or using sideband transitions.\cite{2013Strand,2009Leek,2010Bergeal} While this approach has proven to be useful for few coupled circuit elements, it seems impracticable for larger systems, where it is hard to provide sufficient detunings between all circuit elements. \cite{Makhlin1999} Therefore, one may alternatively use tunable coupling elements such as qubits \cite{Baust2015,2014Baust2, Niskanen2006, Niskanen2007, Hoi2013} or SQUIDs. \cite{vanderPloeg2007, Hime2006, Yin:2013, Pierre2014, Allman2014, xmonexp} One particular example for an interesting application of such actively coupled circuit elements are quantum simulations of bosonic many-body Hamiltonians.\cite{Leib2012,Hartmann2010,PhysRevA.87.053846,Gallemi2015} In such a scenario, the bosonic degrees of freedom can be represented by networks of (possibly nonlinear) superconducting resonators. For this quantum simulator, a tunable coupler would constitute an important control knob. A more general scope of this device is the controllable routing of photonic states on a chip, which is interesting for quantum information as well as quantum simulation experiments.\\
In this work, we experimentally investigate the case of two nearly frequency-degenerate superconducting transmission line resonators coupled by an rf~SQUID in the spirit of Refs.~[\onlinecite{PeropadreTBS,ParaCoupQRes}]. The rf~SQUID, a superconducting loop intersected by a single Josephson junction, acts as a flux-tunable mutual inductance, mediating a flux tunable coupling between the two resonators. Tuning the rf~SQUID by an external flux, we can vary the coupling strength between $\SI{-320}{\mega \hertz}$ and $\SI{+37}{\mega \hertz}$ and observe the corresponding coupled modes in our system. Furthermore, we find that the coupling can be suppressed by approximately four orders of magnitude and, therefore, efficiently be switched off. Finally, the flux-tunable inductance provided by the rf~SQUID gives access to parametric amplification with gains of up to \SI{20}{\deci \bel}. 
Our results are particularly relevant for the realization of scalable arrays of superconducting resonators for quantum information processing and quantum simulation applications. In this context, we emphasize that the rf~SQUID coupler requires only a single Josephson junction and a single control field per coupler.

\section{System Hamiltonian}
\label{Sec:Hamiltonian}
\begin{figure}[h]
  \includegraphics[width=0.5\textwidth]{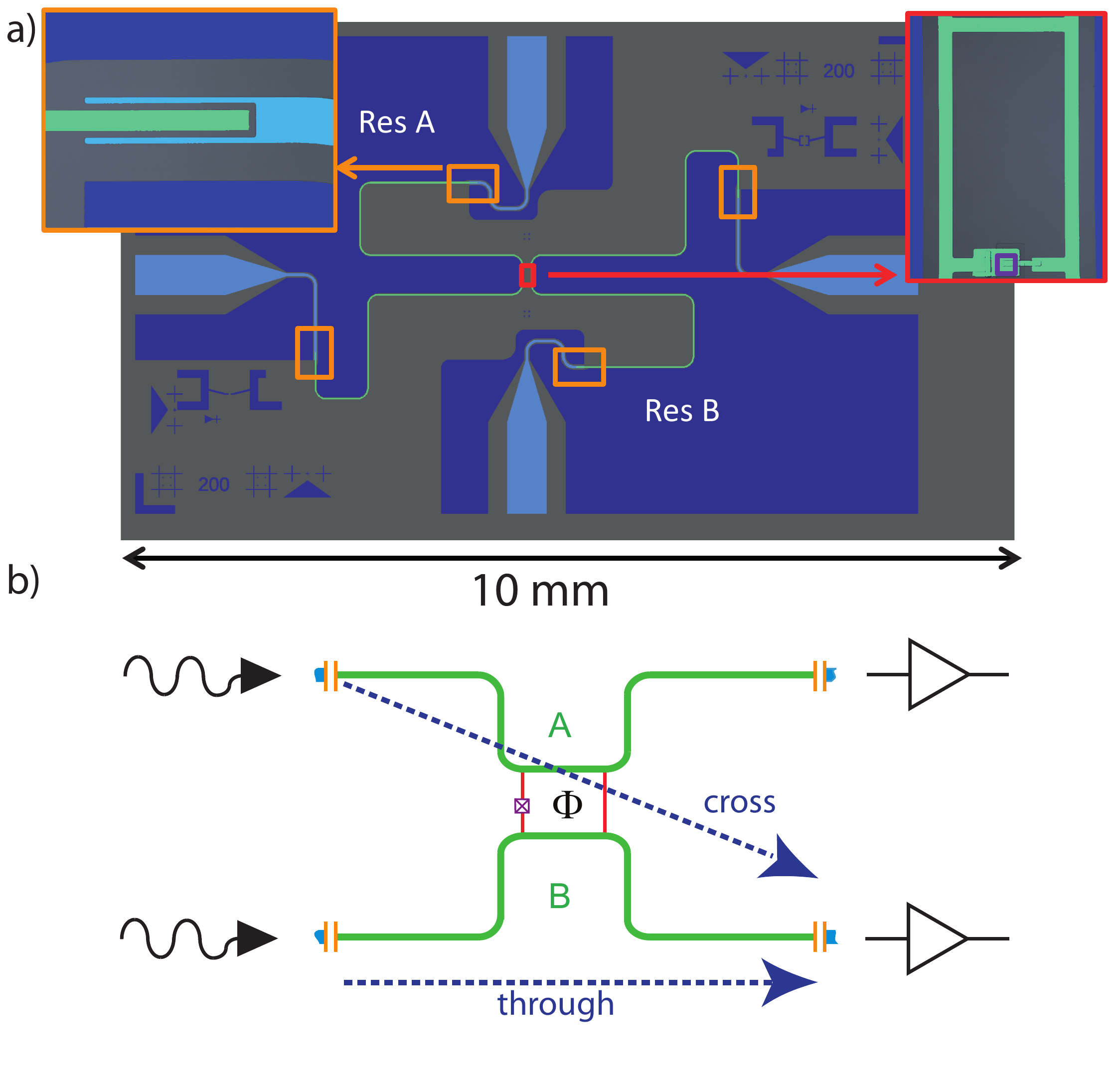}
   \caption{Sample and measurement setup. (a) Sketch of the sample chip. Dark blue: Resonator groundplanes. Green: Resonator center conductors; light blue: Feed line center conductors. The insets are (false color) optical micrographs of the coupling capacitors (orange), the rf~SQUID (red), and the rf~SQUID junction (purple). (b) Sketch of the device measurement setup indicating the attenuated input lines (wiggly arrows), the output lines including cryogenic and room temperature microwave amplifiers (triangular symbols), and possible measurement paths (blue dashed arrows).}
  \label{Wulschner_Fig1}
\end{figure}

An optical micrograph of the sample is shown in Fig.~\ref{Wulschner_Fig1}(a). Our system is comprised of an rf~SQUID galvanically coupled to the center conductor of two coplanar stripline resonators. These resonators, A and B, can be described as quantum harmonic oscillators using the Hamiltonian

\begin{equation}
\label{Hosz}
   H_{\rm{res}}= \hbar \omega_{\rm{A}} a^\dagger a +  \hbar \omega_{\rm{B}} b^\dagger b\,. 
\end{equation}

Here, $\omega_{\rm{A}}$ and $\omega_\text{B}$ are the resonance frequencies and $a^\dagger$, $b^\dagger$, $a$, and $b$ are the bosonic creation and annihilation operators. The effect of the rf~SQUID on the system properties can be modeled in terms of an effective inductance.\cite{medTunfluxQcoup, RFQResCoupling, CoupXTh} The fluxes $\Phi_{\rm{A}}$ and $\Phi_{\rm{B}}$ generated by the resonators in the rf~SQUID give rise to an inductive interaction energy.  
The rf~SQUID consists of a superconducting loop with inductance $L_{\rm{s}}$, which is interrupted by a Josephson junction with critical current $I_{\rm{c}}$. The flux $\Phi$ threading the SQUID loop gives rise to a circulating current 
\begin{equation}
     I_{\rm{s}}(\Phi)=-I_{\rm{c}}\sin(2\pi \Phi/\Phi_{\rm{0}})\,,
		\label{Icir}
\end{equation}

where $\Phi_{\rm{0}}$ is the flux quantum. Here, $\Phi$ is the sum of the externally applied flux $\Phi_{\rm{ext}}$ and the flux generated by $I_{\rm{s}}$,
\begin{equation}
\label{loopflux}
   \Phi=\Phi_{\rm{ext}}+L_{\rm{s}} I_{\rm{s}}(\Phi)\,.		
\end{equation}

Since, in the experiment, the screening parameter $\beta\,{=}\,2\pi L_{\rm{s}} I_{\rm{c}}/\Phi_{\rm{0}}\,{<}\,1$, the dependence of the total flux on the external flux is single-valued. From Eq.~(\ref{Icir}) and Eq.~(\ref{loopflux}), an expression for the effective SQUID inductance is obtained, 

\begin{equation}\label{LRF}
   \frac{1}{L_{\rm{rf}} (\Phi)}=\frac{\partial I_{\rm{s}}}{\partial \Phi_{\rm{ext}}}=-\frac{1}{L_{\rm{s}}}\frac{\beta \cos{(2\pi\frac{\Phi}{\Phi_{\rm{0}}})}}{1+\beta \cos{(2\pi\frac{\Phi}{\Phi_{\rm{0}}})}}\,.
\end{equation}

With this result, we obtain the inductive contribution to the system Hamiltonian,

\begin{equation}\label{Hint}
H_{\rm{ind}}=\frac{(\Phi_{\rm{A}}-\Phi_{\rm{B}})^2}{2 L_{\rm{rf}}(\Phi)}=\frac{\Phi_{\rm{A}}^2+\Phi_{\rm{B}}^2-2\Phi_{\rm{A}}\Phi_{\rm{B}}}{2 L_{\rm{rf}}(\Phi)}\,.
\end{equation}

Calculating the injected fluxes in Eq.~(\ref{Hint}) as described in Ref.~\onlinecite{USCB}, two essential properties of the system become obvious. First, the $\Phi_{\rm{{A,B}}}^2$-terms in Eq.~(\ref{Hint}) lead to dressed resonator frequencies

\begin{align}\label{omegetilde}
        \tilde{\omega}_{\rm{{A,B}}} &= \omega_{\rm{{A,B}}}\sqrt{1-2\frac{L_{\rm{c}}^2}{ L_{\rm{{A,B}}} L_{\rm{rf}}(\Phi)} } \nonumber\\
        &\approx \omega_{\rm{{A,B}}} \left( 1-\frac{L_{\rm{c}}^2  }{  L_{\rm{{A,B}}} L_{\rm{rf}}(\Phi) }  \right) \,,
\end{align}

where $L_{\rm{A,B}}$ is the inductance of the resonators and $L_{\rm{c}}$ the inductance of the segment shared between resonator and rf~SQUID. The second effect of Eq.~(\ref{Hint}), caused by term ${\propto}\,\Phi_{\rm{A}} \Phi_{\rm{B}}$, is a flux dependent coupling 

\begin{equation}\label{gfinal}
g_{\rm{AB}}(\Phi)=-\sqrt{\omega_{\rm{A}} \omega_{\rm{B}}}\frac{  L_{\rm{c}}^2 }{ \sqrt{L_{\rm{A}} L_{\rm{B}}} L_{\rm{rf}}(\Phi)} 
\end{equation}

between the resonators. Due to their vicinity on the chip, there is a also flux independent direct inductive coupling component $g_{\rm{0}}$ between the resonators. Thus the total coupling reads

\begin{equation}\label{netg}
g(\Phi)=g_{\rm{AB}}(\Phi)+g_{\rm{0}}.
\end{equation}

Equation~(\ref{LRF}) shows that $g_{\rm{AB}}(\Phi)$ can be positive or negative depending on the applied flux. By applying a suitable flux, the rf~SQUID mediated coupling compensates the direct inductive coupling. In this way one can turn on and off the net coupling between the resonators. After a rotating wave approximation the full Hamiltonian reads
\begin{equation}
\label{fullhamiltonian}
H= 
\hbar
\begin{pmatrix}
a^\dagger & b^\dagger \\
\end{pmatrix}
\begin{pmatrix}
\tilde{\omega}_{\rm{A}} & g(\Phi) \\
g(\Phi) &\tilde{\omega}_{\rm{B}} \\
\end{pmatrix}
\begin{pmatrix}
a \\
b \\
\end{pmatrix}.
\end{equation}

The eigenvalues of Eq.~(\ref{fullhamiltonian}) correspond to the new eigenfrequencies
\begin{equation}
\Omega_{\rm{1,2}}=\frac{\tilde{\omega}_{\rm{A}}+\tilde{\omega}_{\rm{B}}}{2} 
\pm  \sqrt{g(\Phi)^2+\frac{(\tilde{\omega}_{\rm{A}}-\tilde{\omega}_{\rm{B}})^2}{4}}\, .
\label{eigenvalues}
\end{equation}

\section{Sample and Measurement Setup}

\label{Sec:Setup}

\begin{figure*}[htb]
   \includegraphics[width=0.95\textwidth]{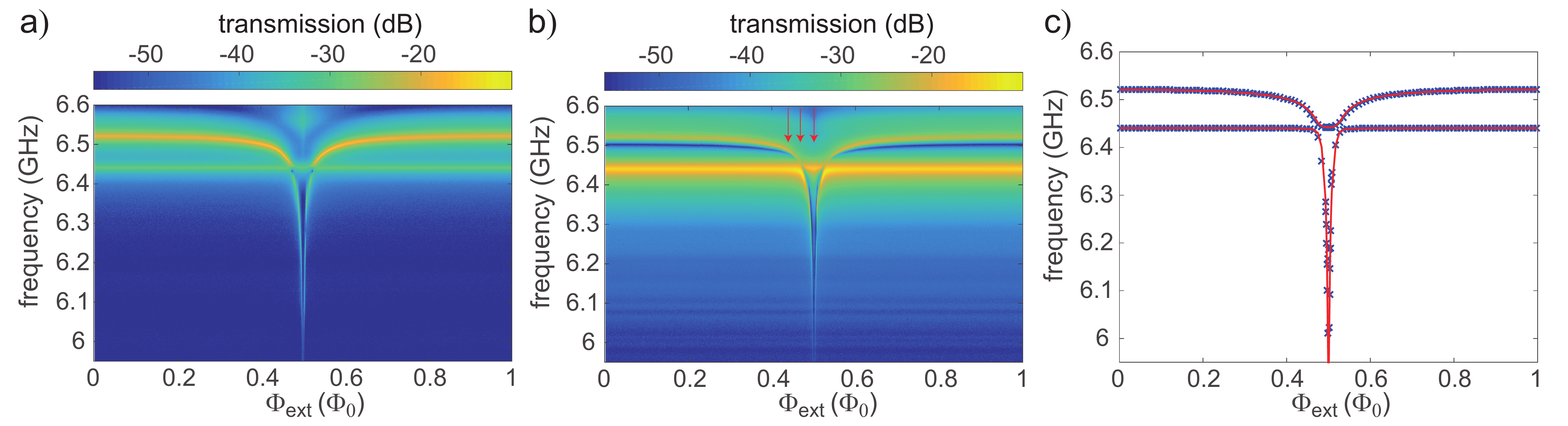}
   \caption{Through transmission magnitude (color coded) as a function of probe frequency and externally applied flux for (a) resonator A and (b) resonator B. The red arrows mark the flux values, for which transmission vs. frequency cuts are shown in Fig.~\ref{fluxcuts}. (c) Fit (red line) of Eq.~(\ref{eigenvalues}) to the extracted center frequencies (crosses). The modes are take from the trough measurements, where they are more pronounced. This is especially necessary when the coupling is much smaller then the detuning of the resonators.}
  \label{eigenmodes}
\end{figure*} 
Figure~\ref{Wulschner_Fig1}(a) shows the layout of the sample chip. In the resonator design, we omit the second groundplane to reduce the direct geometric coupling between the two resonators. The rf~SQUID is galvanically connected to both center strips of the resonators over a length of $\SI{200}{\micro \meter}$. The sample is fabricated as follows. First, a $\SI{100}{\nano \meter}$ thick niobium layer is sputter deposited onto a $\SI{250}{\micro \meter}$ thick, thermally oxidized silicon wafer. The resonators and the SQUID loop are patterned using optical lithography and reactive ion etching. The Josephson junction of the SQUID is fabricated in a $\rm{Nb/AlO_x/Nb}$ trilayer process. The resonators have a characteristic impedance of $Z_0\,{=}\,\SI{64}{\ohm}$ and the resonance frequencies $\omega_{\rm{A}}/2\pi\,{=}\,\SI{6.461}{\giga\hertz}$ and $\omega_{\rm{B}}/2\pi\,{=}\,\SI{6.482}{\giga\hertz}$. The SQUID loop has dimensions of $\SI{200}{\micro \meter} \,{\times}\, \SI{100}{\micro \meter}$ and a screening parameter $\beta\,{=}\, 0.934$ to maximize the coupling according to Eq.~(\ref{LRF}) while keeping the SQUID mono\-stable. The sample is mounted inside a gold plated copper box, which is attached to the base temperature stage of a dilution refrigerator operating at $\SI{26}{\milli \kelvin}$. A superconducting solenoid attached to the top of the sample box is used to generate the external flux applied to the rf~SQUID. As depicted in Fig.~\ref{Wulschner_Fig1}(b), one port of each resonator is connected to an attenuated input microwave line, whereas the remaining ports are connected to output lines containing microwave amplifiers.

\section{Resonator spectroscopy}
\begin{figure}[htp]
   \includegraphics[width=0.4\textwidth]{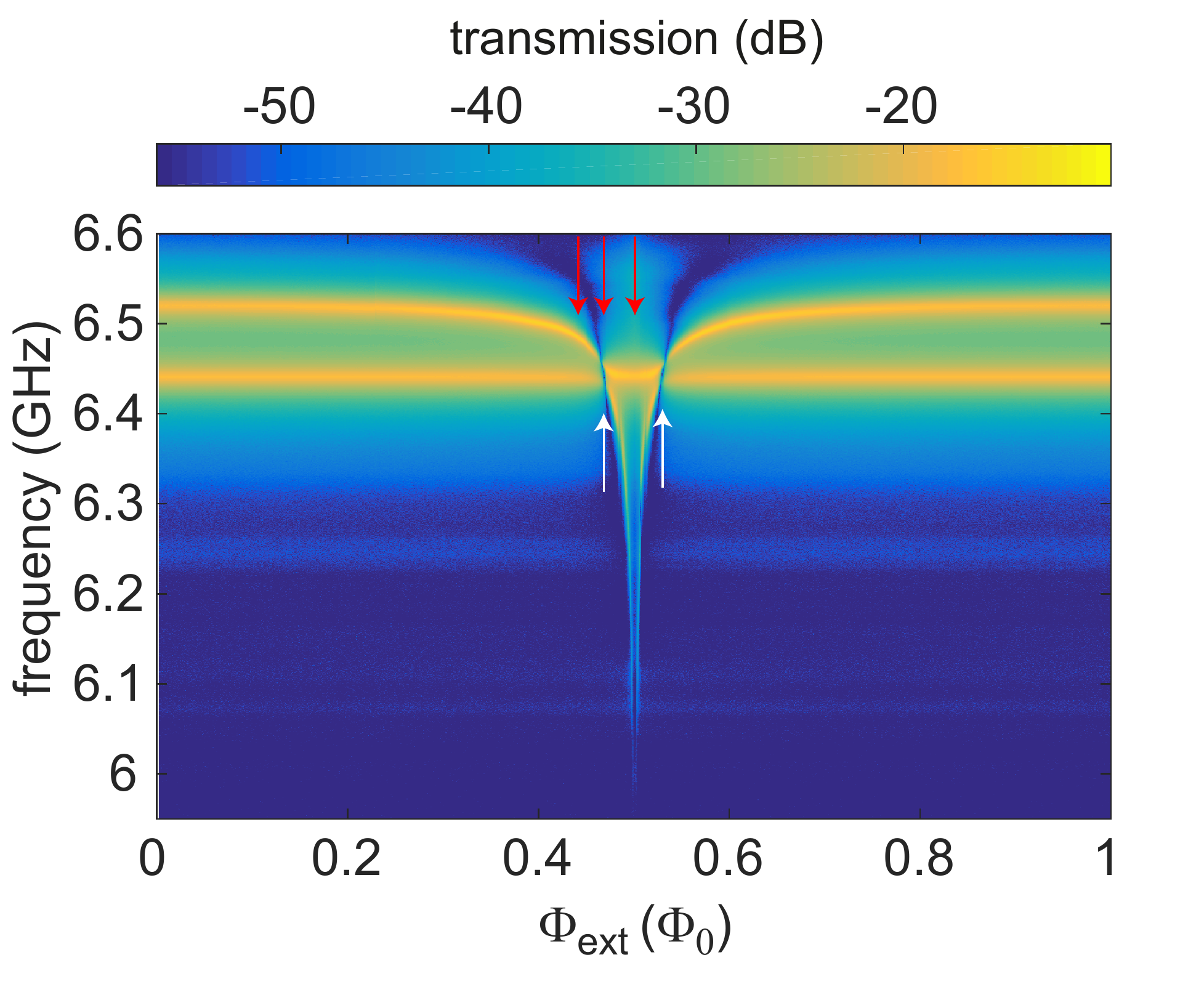}
   \caption{Cross-transmission from resonator A to resonator B as a function of the applied flux. Near $\Phi_{\rm{ext}}\,{=}\,0.468 \, \Phi_{\rm{0}}$ and $\Phi_{\rm{ext}}\,{=}\,0.532 \, \Phi_{\rm{0}}$ (white arrows),  the resonators decouple and signal transmission is blocked. The red arrows mark the flux values, for which transmission vs. frequency cuts are shown in Fig.~\ref{fluxcuts}.}
  \label{crossmeasurment}
\end{figure}

We first extract the properties of the rf~SQUID coupler from transmission measurements through the individual resonators. As indicated in Fig.~1, we call this type of measurement a ``through-measurement''. In contrast, in the ``cross-measurements'' we inject a signal into one of the resonators and probe the output of the other one. In these measurements, the applied microwave power corresponds to an average photon number of about one in the resonators. In Fig.~\ref{eigenmodes}(a) and Fig.~\ref{eigenmodes}(b), the through measurements of resonator A and B are shown depending on the applied flux $\Phi_{\rm{ext}}$. According to Eq.~(\ref{LRF}) and Eq.~(\ref{Hint}), the modulation of the resonator modes due to the presence of the rf~SQUID is $\Phi_{\rm{0}}$-periodic and symmetric with respect to $\Phi_{\rm{ext}}\,{=}\, \Phi_{\rm{0}}/2$. The two modes of Eq.~(\ref{eigenvalues}) manifest themselves as two resonances in the spectroscopy data. As expected, we observe a flux dependent mode distance, caused by the flux tunable mutual inductance of the rf~SQUID. For most flux values, one observes two resonance peaks independent of the chosen input and output port. However near $\Phi_{\rm{ext}} \, {=}\, 0.468\, \Phi_{\rm{0}}$ and $\Phi_{\rm{ext}} \, {=}\,0.532 \, \Phi_{\rm{0}}$, only a single peak is present in the through measurements and cross transmission is strongly suppressed (see Fig.~\ref{crossmeasurment}). These are the points where the SQUID-mediated coupling compensates the direct inductive coupling resulting in a vanishing total coupling and, hence, completely decoupled resonators. Note that for $g(\Phi)\,{=}\, 0$, the Hamiltonian of Eq.~(\ref{fullhamiltonian}) becomes diagonal and each of the two modes reflects the excitation of one of the resonators. In Fig.~\ref{eigenmodes}(c), the center frequencies of the normal modes $\Omega_{\rm{1,2}}$ derived from the data in Fig.~\ref{eigenmodes}(a) and Fig.~\ref{eigenmodes}(b) are plotted along with a fit using Eq.~(\ref{eigenvalues}). 
 \begin{figure*}[htb]
   \includegraphics[width=0.95\textwidth]{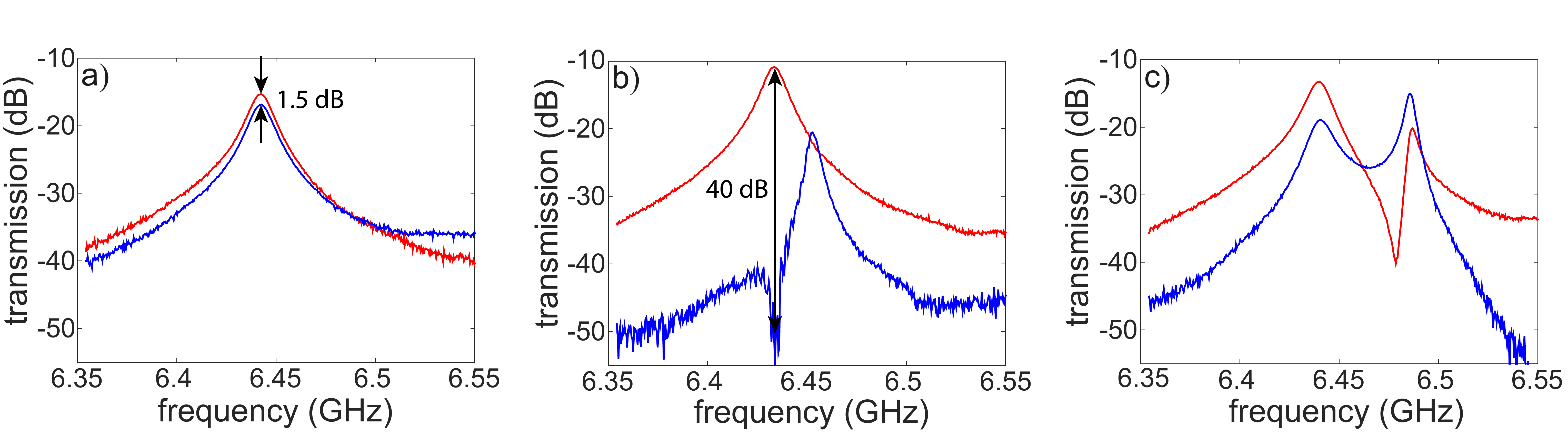}
   \caption{Through- (red) and cross-measurements (blue), obtained for three different flux values. (a) $\Phi_{\rm{ext}}\,{=}\,\Phi_0/2$,$|g|\,{\gg}\,|\Delta|$: strongly coupled regime, only the mode $\Omega_1$ is shown. (b) $\Phi_{\rm{ext}}\,{=}\,0.468\Phi_0/2$, $|g|\,{\simeq}\, 0$: decoupling point. (c) $\Phi_{\rm{ext}}\,{=}\,0.439\Phi_0/2$, $|g|\,{\simeq}\,\Delta$.}
  \label{fluxcuts}
\end{figure*} Upon closer inspection, we find a minimum distance on the order of $\SI{20}{\mega\hertz}$ between these modes. This finite gap is caused by a small detuning  $\Delta\,{=}\,\omega_{\rm{A}}\,{-}\,\omega_{\rm{B}} \,{=}\, 2 \pi \,{\times}\, \SI{21.3}{\mega\hertz}$ of the resonators. We also observe different decay rates of the resonators, which we extract from the through measurements of both resonators at a decoupling point. Lorentzian fits lead to $\gamma_{\rm{A}}/ 2 \pi \,{=}\, \SI{3.6}{\mega\hertz}$ and  $\gamma_{\rm{B}}/ 2 \pi\,{=}\, \SI{6.1}{\mega\hertz}$. The fact that resonator A has smaller width at half maximum and a and a slightly higher eigenfrequency than resonator B, indicates a smaller effective coupling capacitance,\cite{Goeppl2008} which we attribute to fabrication or sample contacting imperfections. Therefore, we assume $L_{\rm{A}}/L_{\rm{B}}\,{\approx}\, 1$. Furthermore, we define the fitting parameter $g_{\rm{0}}\,{=}\,\sqrt{\omega_{\rm{A}} \omega_{\rm{B}}  }L_{\rm{c}}^2 /(\sqrt{L_{\rm{A}} L_{\rm{B}}} L_{\rm{s}})$. In this way, the rf~SQUID coupling reads $g_{\rm{AB}}\,{=}\, g_{\rm{0}} \beta \cos{(2\pi\frac{\Phi}{\Phi_{\rm{0}}})}/(1{+}\beta \cos{(2\pi\frac{\Phi}{\Phi_{\rm{0}}})})$. Fitting the data, we obtain $g_{\rm{0}}/ 2 \pi\,{=}\, \SI{29.0}{\mega\hertz}$ and $g_{\rm{0}}/ 2 \pi\,{=}\, \SI{20.4}{\mega\hertz}$. From the mode distance we find $g(\Phi)/ 2 \pi$ ranging between $\SI{+37}{\mega\hertz}$ and $\SI{-320}{\mega\hertz}$. Beyond this value, the lower mode becomes too broad due to its steep flux dependence. \\
Next, we analyze the properties of our device in the coupled and decoupled state in more detail. Because of the small detuning of the resonators, the coupled modes are not necessarily symmetric and antisymmetric superpositions of the uncoupled modes. This is also seen in the spectroscopy of the single resonators (see\,Fig.~\ref{eigenmodes}), where the modes have different intensities. The mode mixing can be estimated from the eigenvectors of the Hamiltonian in Eq.~(\ref{fullhamiltonian}). For $g/2\pi\,{=}\, \SI{+37}{\mega\hertz}$ and $g/2\pi\,{=}\,\SI{-320}{\mega\hertz}$ we obtain the mixing ratios 63:37 and 52:48, respectively. Hence, in the latter case, our sample satisfies the condition $|g|\,{\gg}\,|\Delta|$, where the detuning becomes insignificant. In the decoupled case near $\Phi_{\rm{ext}}\,{=}\,0.468 \, \Phi_{\rm{0}}$ and $\Phi_{\rm{ext}}\,{=}\,0.532 \, \Phi_{\rm{0}}$, the off-diagonal elements in the Hamiltonian of Eq.~(\ref{fullhamiltonian}) vanish and the modes are pure excitations of resonator A or B.\\
In this situation, it is particularly instructive to examine the cross-transmission spectra such as the one shown in Fig.~\ref{crossmeasurment}, where resonator A is driven and resonator B is probed. Here, we clearly see that in two narrow regions around $\Phi_{\rm{ext}}\,{=}\, 0.468 \Phi_{\rm{0}}$ and $\Phi_{\rm{ext}}\,{=}\, 0.532 \Phi_{\rm{0}}$, where the net coupling $g(\Phi)$ approaches zero, the microwave transmission between the resonators is blocked. We gain further insight by comparing the through- and cross-transmission spectra in Fig.~\ref{fluxcuts}(a) and Fig.~\ref{fluxcuts}(b). For $|g(\Phi)| \gg |\Delta|$ [see Fig.~\ref{fluxcuts}(a)], both measurements exhibit similar peak heights. Since both measurements use the same output line, we relate the small difference of approximately \SI{1.5}{\deci\bel} mainly to the slightly different losses in the input lines. For $g(\Phi) \,{\approx}\, 0$, however, the cross-transmission is suppressed by \SI{40}{\deci \bel} on resonance as shown in Fig.~\ref{fluxcuts}(b), corresponding to a relative transmission change of \SI{38.5}{\deci \bel}. This result confirms that we can sufficiently compensate the direct inductive coupling with the tunable SQUID-mediated coupling. Finally, in Fig.~\ref{fluxcuts}(c) we show the transmission for a flux value, where $g(\Phi)$ and $\Delta$ are comparable. In the through measurement, the detuning manifests itself in the form of unequal peak heights and an anti-resonance dip, which is not centered between the resonance peaks.\cite{Wahl1999,Antires}

\section{Parametric amplification}

\begin{figure}
  \includegraphics[width=0.4\textwidth]{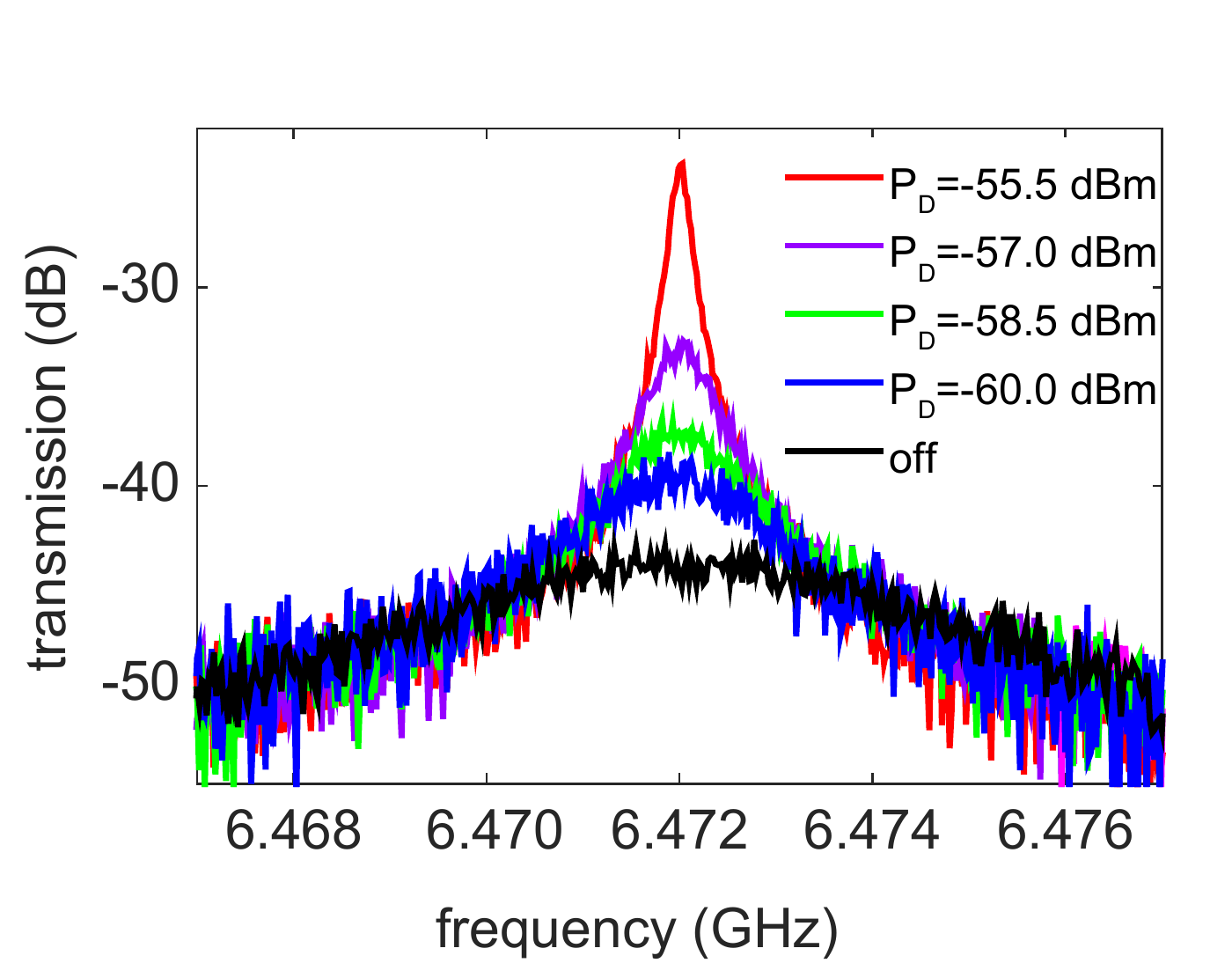}
   \caption{Mode $\Omega_{\rm{1}}$ for $\Phi_\text{ext}\,{=}\,0.450 \,\Phi_0$ with an additional drive tone $\omega_{\rm{D}}/2\pi\,{=}\,\SI{12.944}{\giga \hertz}$ for different drive powers $\rm{P_D}$. The drive power refers to the power at sample input.}
  \label{paramp}
\end{figure}

So far, we have treated the rf~SQUID as a flux-tunable coupler between the two resonators. However, we can also interpret it as a flux-tunable inductance, parametrically modifying the eigenfrequency of the modes. Then, a harmonic flux drive applied in addition to the static flux allows for the study of parametric effects. Our experiments in this direction are inspired by analog experiments on Josephson parametric amplifier \cite{Zhong,2012Menzel,JPA} and converter.\cite{JPC} In order to operate our device as a flux-driven amplifier, we apply an additional drive tone at $\omega_{\rm{D}}\,{=}\,2\Omega_{\rm{1}}$ to the input line. The drive tone periodically modulates the flux threading the SQUID loop and therefore the mode frequency, leading to parametric amplification. To characterize the performance of the resulting parametric amplifier, we calculate the power gain (G) as well as the bandwidth characterized by the full width half maximum $\Delta\Omega$ of the amplified signals. Figure~\ref{paramp} shows the transmission in the vicinity of $\Omega_{\rm{1}}/2\pi\,{=}\,\SI{6.472}{\giga\hertz}$ and $\Phi_\text{ext}\,{=}\,0.450 \,\Phi_0$ for different values of the drive power. While the gain is increasing for higher drive strength, the bandwidth decreases as expected. For a nondegenerate (phase-insensitive) gain of $G\,{=}\,\SI{20}{\deci \bel}$, the gain-bandwidth product is $G\Delta\Omega/2\pi=\SI{21.5}{\mega\hertz}$. Checking the theoretical relation\cite{2006JPATheo} $\sqrt{G}\Delta \Omega(G)\,{=}\, \rm{const.}$, we find a maximum deviation of 2.5 similar to other experiments.\cite{arrayJPA}

\section{Conclusions}
In conclusion, we present a flux-tunable coupling between two superconducting resonators based on an rf~SQUID. Spectroscopically, we measure negative and positive couplings ranging from $\SI{-320}{\mega \hertz}$ to $\SI{37}{\mega \hertz}$. Furthermore, the observed suppression of the cross-transmission of up to $\SI{38.5}{\deci \bel}$ proves the ability to effectively turn off the coupling and is a significant improvement compared to previous work.\cite{Baust2015,2014Baust2} With this performance, our coupler is considered a useful tool both for quantum computation, where a controlled nearest neighbor interaction is required and our coupler could be used to route information in a controllable way on a chip and for quantum simulation experiments,\cite{PhysRevA.87.053846, Houck2012, Raftery2013, Hartmann2010, Leib2012} where it could be useful not only to change the amplitude of the coupling constant but also its sign.\cite{Gallemi2015} Finally, we investigate parametric amplification of the device and observe a nondegenerate gain as large as $\SI{20}{\deci \bel}$.

The authors acknowledge support from the German Research Foundation through SFB 631 and FE 1564/1-1, EU projects CCQED, PROMISCE and SCALEQIT, the doctorate program ExQM of the Elite Network of Bavaria, Spanish MINECO Projects FIS2012-33022 and FIS2012-36673-C03-02, CAM Research Network QUITEMAD+, Basque Government IT472-10 and UPV/EHU UFI 11/55. E.S. acknowledges support from a TUM August-Wilhelm Scheer Visiting Professorship and hospitality of Walther-Mei{\ss}ner-Institut and TUM Institute for Advanced Study.


\begin{thebibliography}{36}%
\makeatletter
\providecommand \@ifxundefined [1]{%
 \@ifx{#1\undefined}
}%
\providecommand \@ifnum [1]{%
 \ifnum #1\expandafter \@firstoftwo
 \else \expandafter \@secondoftwo
 \fi
}%
\providecommand \@ifx [1]{%
 \ifx #1\expandafter \@firstoftwo
 \else \expandafter \@secondoftwo
 \fi
}%
\providecommand \natexlab [1]{#1}%
\providecommand \enquote  [1]{``#1''}%
\providecommand \bibnamefont  [1]{#1}%
\providecommand \bibfnamefont [1]{#1}%
\providecommand \citenamefont [1]{#1}%
\providecommand \href@noop [0]{\@secondoftwo}%
\providecommand \href [0]{\begingroup \@sanitize@url \@href}%
\providecommand \@href[1]{\@@startlink{#1}\@@href}%
\providecommand \@@href[1]{\endgroup#1\@@endlink}%
\providecommand \@sanitize@url [0]{\catcode `\\12\catcode `\$12\catcode
  `\&12\catcode `\#12\catcode `\^12\catcode `\_12\catcode `\%12\relax}%
\providecommand \@@startlink[1]{}%
\providecommand \@@endlink[0]{}%
\providecommand \url  [0]{\begingroup\@sanitize@url \@url }%
\providecommand \@url [1]{\endgroup\@href {#1}{\urlprefix }}%
\providecommand \urlprefix  [0]{URL }%
\providecommand \Eprint [0]{\href }%
\providecommand \doibase [0]{http://dx.doi.org/}%
\providecommand \selectlanguage [0]{\@gobble}%
\providecommand \bibinfo  [0]{\@secondoftwo}%
\providecommand \bibfield  [0]{\@secondoftwo}%
\providecommand \translation [1]{[#1]}%
\providecommand \BibitemOpen [0]{}%
\providecommand \bibitemStop [0]{}%
\providecommand \bibitemNoStop [0]{.\EOS\space}%
\providecommand \EOS [0]{\spacefactor3000\relax}%
\providecommand \BibitemShut  [1]{\csname bibitem#1\endcsname}%
\let\auto@bib@innerbib\@empty
%</preamble>
\bibitem [{\citenamefont {Strand}\ \emph {et~al.}(2013)\citenamefont {Strand},
  \citenamefont {Ware}, \citenamefont {Beaudoin}, \citenamefont {Ohki},
  \citenamefont {Johnson}, \citenamefont {Blais},\ and\ \citenamefont
  {Plourde}}]{2013Strand}%
  \BibitemOpen
  \bibfield  {author} {\bibinfo {author} {\bibfnamefont {J.~D.}\ \bibnamefont
  {Strand}}, \bibinfo {author} {\bibfnamefont {M.}~\bibnamefont {Ware}},
  \bibinfo {author} {\bibfnamefont {F.}~\bibnamefont {Beaudoin}}, \bibinfo
  {author} {\bibfnamefont {T.~A.}\ \bibnamefont {Ohki}}, \bibinfo {author}
  {\bibfnamefont {B.~R.}\ \bibnamefont {Johnson}}, \bibinfo {author}
  {\bibfnamefont {A.}~\bibnamefont {Blais}}, \ and\ \bibinfo {author}
  {\bibfnamefont {B.~L.~T.}\ \bibnamefont {Plourde}},\ }\href {\doibase
  10.1103/PhysRevB.87.220505} {\bibfield  {journal} {\bibinfo  {journal} {Phys.
  Rev. B}\ }\textbf {\bibinfo {volume} {87}},\ \bibinfo {pages} {220505}
  (\bibinfo {year} {2013})}\BibitemShut {NoStop}%
\bibitem [{\citenamefont {Leek}\ \emph {et~al.}(2009)\citenamefont {Leek},
  \citenamefont {Filipp}, \citenamefont {Maurer}, \citenamefont {Baur},
  \citenamefont {Bianchetti}, \citenamefont {Fink}, \citenamefont {G\"oppl},
  \citenamefont {Steffen},\ and\ \citenamefont {Wallraff}}]{2009Leek}%
  \BibitemOpen
  \bibfield  {author} {\bibinfo {author} {\bibfnamefont {P.~J.}\ \bibnamefont
  {Leek}}, \bibinfo {author} {\bibfnamefont {S.}~\bibnamefont {Filipp}},
  \bibinfo {author} {\bibfnamefont {P.}~\bibnamefont {Maurer}}, \bibinfo
  {author} {\bibfnamefont {M.}~\bibnamefont {Baur}}, \bibinfo {author}
  {\bibfnamefont {R.}~\bibnamefont {Bianchetti}}, \bibinfo {author}
  {\bibfnamefont {J.~M.}\ \bibnamefont {Fink}}, \bibinfo {author}
  {\bibfnamefont {M.}~\bibnamefont {G\"oppl}}, \bibinfo {author} {\bibfnamefont
  {L.}~\bibnamefont {Steffen}}, \ and\ \bibinfo {author} {\bibfnamefont
  {A.}~\bibnamefont {Wallraff}},\ }\href {\doibase 10.1103/PhysRevB.79.180511}
  {\bibfield  {journal} {\bibinfo  {journal} {Phys. Rev. B}\ }\textbf {\bibinfo
  {volume} {79}},\ \bibinfo {pages} {180511} (\bibinfo {year}
  {2009})}\BibitemShut {NoStop}%
\bibitem [{\citenamefont {{Bergeal}}\ \emph {et~al.}(2010)\citenamefont
  {{Bergeal}}, \citenamefont {{Vijay}}, \citenamefont {{Manucharyan}},
  \citenamefont {{Siddiqi}}, \citenamefont {{Schoelkopf}}, \citenamefont
  {{Girvin}},\ and\ \citenamefont {{Devoret}}}]{2010Bergeal}%
  \BibitemOpen
  \bibfield  {author} {\bibinfo {author} {\bibfnamefont {N.}~\bibnamefont
  {{Bergeal}}}, \bibinfo {author} {\bibfnamefont {R.}~\bibnamefont {{Vijay}}},
  \bibinfo {author} {\bibfnamefont {V.~E.}\ \bibnamefont {{Manucharyan}}},
  \bibinfo {author} {\bibfnamefont {I.}~\bibnamefont {{Siddiqi}}}, \bibinfo
  {author} {\bibfnamefont {R.~J.}\ \bibnamefont {{Schoelkopf}}}, \bibinfo
  {author} {\bibfnamefont {S.~M.}\ \bibnamefont {{Girvin}}}, \ and\ \bibinfo
  {author} {\bibfnamefont {M.~H.}\ \bibnamefont {{Devoret}}},\ }\href
  {http://dx.doi.org/10.1038/nphys1516} {\bibfield  {journal} {\bibinfo
  {journal} {Nat Phys}\ }\textbf {\bibinfo {volume} {6}},\ \bibinfo {pages}
  {296} (\bibinfo {year} {2010})}\BibitemShut {NoStop}%
\bibitem [{\citenamefont {{Makhlin}}\ \emph {et~al.}(1999)\citenamefont
  {{Makhlin}}, \citenamefont {{Sch{\"o}n}},\ and\ \citenamefont
  {{Shnirman}}}]{Makhlin1999}%
  \BibitemOpen
  \bibfield  {author} {\bibinfo {author} {\bibfnamefont {Y.}~\bibnamefont
  {{Makhlin}}}, \bibinfo {author} {\bibfnamefont {G.}~\bibnamefont
  {{Sch{\"o}n}}}, \ and\ \bibinfo {author} {\bibfnamefont {A.}~\bibnamefont
  {{Shnirman}}},\ }\href {http://dx.doi.org/10.1038/18613} {\bibfield
  {journal} {\bibinfo  {journal} {Nature}\ }\textbf {\bibinfo {volume} {398}},\
  \bibinfo {pages} {305} (\bibinfo {year} {1999})}\BibitemShut {NoStop}%
\bibitem [{\citenamefont {Baust}\ \emph {et~al.}(2015)\citenamefont {Baust},
  \citenamefont {Hoffmann}, \citenamefont {Haeberlein}, \citenamefont
  {Schwarz}, \citenamefont {Eder}, \citenamefont {Goetz}, \citenamefont
  {Wulschner}, \citenamefont {Xie}, \citenamefont {Zhong}, \citenamefont
  {Quijandr\'{i}a}, \citenamefont {Peropadre}, \citenamefont {Zueco},
  \citenamefont {Garc\'{i}a~Ripoll}, \citenamefont {Solano}, \citenamefont
  {Fedorov}, \citenamefont {Menzel}, \citenamefont {Deppe}, \citenamefont
  {Marx},\ and\ \citenamefont {Gross}}]{Baust2015}%
  \BibitemOpen
  \bibfield  {author} {\bibinfo {author} {\bibfnamefont {A.}~\bibnamefont
  {Baust}}, \bibinfo {author} {\bibfnamefont {E.}~\bibnamefont {Hoffmann}},
  \bibinfo {author} {\bibfnamefont {M.}~\bibnamefont {Haeberlein}}, \bibinfo
  {author} {\bibfnamefont {M.~J.}\ \bibnamefont {Schwarz}}, \bibinfo {author}
  {\bibfnamefont {P.}~\bibnamefont {Eder}}, \bibinfo {author} {\bibfnamefont
  {J.}~\bibnamefont {Goetz}}, \bibinfo {author} {\bibfnamefont
  {F.}~\bibnamefont {Wulschner}}, \bibinfo {author} {\bibfnamefont
  {E.}~\bibnamefont {Xie}}, \bibinfo {author} {\bibfnamefont {L.}~\bibnamefont
  {Zhong}}, \bibinfo {author} {\bibfnamefont {F.}~\bibnamefont
  {Quijandr\'{i}a}}, \bibinfo {author} {\bibfnamefont {B.}~\bibnamefont
  {Peropadre}}, \bibinfo {author} {\bibfnamefont {D.}~\bibnamefont {Zueco}},
  \bibinfo {author} {\bibfnamefont {J.-J.}\ \bibnamefont {Garc\'{i}a~Ripoll}},
  \bibinfo {author} {\bibfnamefont {E.}~\bibnamefont {Solano}}, \bibinfo
  {author} {\bibfnamefont {K.}~\bibnamefont {Fedorov}}, \bibinfo {author}
  {\bibfnamefont {E.~P.}\ \bibnamefont {Menzel}}, \bibinfo {author}
  {\bibfnamefont {F.}~\bibnamefont {Deppe}}, \bibinfo {author} {\bibfnamefont
  {A.}~\bibnamefont {Marx}}, \ and\ \bibinfo {author} {\bibfnamefont
  {R.}~\bibnamefont {Gross}},\ }\href {\doibase 10.1103/PhysRevB.91.014515}
  {\bibfield  {journal} {\bibinfo  {journal} {Phys. Rev. B}\ }\textbf {\bibinfo
  {volume} {91}},\ \bibinfo {pages} {014515} (\bibinfo {year}
  {2015})}\BibitemShut {NoStop}%
\bibitem [{\citenamefont {{Baust}}\ \emph {et~al.}(2014)\citenamefont
  {{Baust}}, \citenamefont {{Hoffmann}}, \citenamefont {{Haeberlein}},
  \citenamefont {{Schwarz}}, \citenamefont {{Eder}}, \citenamefont {{Goetz}},
  \citenamefont {{Wulschner}}, \citenamefont {{Xie}}, \citenamefont {{Zhong}},
  \citenamefont {{Quijandria}}, \citenamefont {{Zueco}}, \citenamefont {{Garcia
  Ripoll}}, \citenamefont {{Garcia-Alvarez}}, \citenamefont {{Romero}},
  \citenamefont {{Solano}}, \citenamefont {{Fedorov}}, \citenamefont
  {{Menzel}}, \citenamefont {{Deppe}}, \citenamefont {{Marx}},\ and\
  \citenamefont {{Gross}}}]{2014Baust2}%
  \BibitemOpen
  \bibfield  {author} {\bibinfo {author} {\bibfnamefont {A.}~\bibnamefont
  {{Baust}}}, \bibinfo {author} {\bibfnamefont {E.}~\bibnamefont {{Hoffmann}}},
  \bibinfo {author} {\bibfnamefont {M.}~\bibnamefont {{Haeberlein}}}, \bibinfo
  {author} {\bibfnamefont {M.~J.}\ \bibnamefont {{Schwarz}}}, \bibinfo {author}
  {\bibfnamefont {P.}~\bibnamefont {{Eder}}}, \bibinfo {author} {\bibfnamefont
  {J.}~\bibnamefont {{Goetz}}}, \bibinfo {author} {\bibfnamefont
  {F.}~\bibnamefont {{Wulschner}}}, \bibinfo {author} {\bibfnamefont
  {E.}~\bibnamefont {{Xie}}}, \bibinfo {author} {\bibfnamefont
  {L.}~\bibnamefont {{Zhong}}}, \bibinfo {author} {\bibfnamefont
  {F.}~\bibnamefont {{Quijandria}}}, \bibinfo {author} {\bibfnamefont
  {D.}~\bibnamefont {{Zueco}}}, \bibinfo {author} {\bibfnamefont {J.-J.}\
  \bibnamefont {{Garcia Ripoll}}}, \bibinfo {author} {\bibfnamefont
  {L.}~\bibnamefont {{Garcia-Alvarez}}}, \bibinfo {author} {\bibfnamefont
  {G.}~\bibnamefont {{Romero}}}, \bibinfo {author} {\bibfnamefont
  {E.}~\bibnamefont {{Solano}}}, \bibinfo {author} {\bibfnamefont {K.~G.}\
  \bibnamefont {{Fedorov}}}, \bibinfo {author} {\bibfnamefont {E.~P.}\
  \bibnamefont {{Menzel}}}, \bibinfo {author} {\bibfnamefont {F.}~\bibnamefont
  {{Deppe}}}, \bibinfo {author} {\bibfnamefont {A.}~\bibnamefont {{Marx}}}, \
  and\ \bibinfo {author} {\bibfnamefont {R.}~\bibnamefont {{Gross}}},\
  }\href@noop {} {\bibfield  {journal} {\bibinfo  {journal} {ArXiv e-prints}\ }
  (\bibinfo {year} {2014})},\ \Eprint {http://arxiv.org/abs/1412.7372}
  {arXiv:1412.7372} \BibitemShut {NoStop}%
\bibitem [{\citenamefont {Niskanen}\ \emph {et~al.}(2006)\citenamefont
  {Niskanen}, \citenamefont {Nakamura},\ and\ \citenamefont
  {Tsai}}]{Niskanen2006}%
  \BibitemOpen
  \bibfield  {author} {\bibinfo {author} {\bibfnamefont {A.~O.}\ \bibnamefont
  {Niskanen}}, \bibinfo {author} {\bibfnamefont {Y.}~\bibnamefont {Nakamura}},
  \ and\ \bibinfo {author} {\bibfnamefont {J.-S.}\ \bibnamefont {Tsai}},\
  }\href {\doibase 10.1103/PhysRevB.73.094506} {\bibfield  {journal} {\bibinfo
  {journal} {Phys. Rev. B}\ }\textbf {\bibinfo {volume} {73}},\ \bibinfo
  {pages} {094506} (\bibinfo {year} {2006})}\BibitemShut {NoStop}%
\bibitem [{\citenamefont {Niskanen}\ \emph {et~al.}(2007)\citenamefont
  {Niskanen}, \citenamefont {Harrabi}, \citenamefont {Yoshihara}, \citenamefont
  {Nakamura}, \citenamefont {Lloyd},\ and\ \citenamefont
  {Tsai}}]{Niskanen2007}%
  \BibitemOpen
  \bibfield  {author} {\bibinfo {author} {\bibfnamefont {A.~O.}\ \bibnamefont
  {Niskanen}}, \bibinfo {author} {\bibfnamefont {K.}~\bibnamefont {Harrabi}},
  \bibinfo {author} {\bibfnamefont {F.}~\bibnamefont {Yoshihara}}, \bibinfo
  {author} {\bibfnamefont {Y.}~\bibnamefont {Nakamura}}, \bibinfo {author}
  {\bibfnamefont {S.}~\bibnamefont {Lloyd}}, \ and\ \bibinfo {author}
  {\bibfnamefont {J.~S.}\ \bibnamefont {Tsai}},\ }\href {\doibase
  10.1126/science.1141324} {\bibfield  {journal} {\bibinfo  {journal}
  {Science}\ }\textbf {\bibinfo {volume} {316}},\ \bibinfo {pages} {723}
  (\bibinfo {year} {2007})}\BibitemShut {NoStop}%
\bibitem [{\citenamefont {Hoi}\ \emph {et~al.}(2013)\citenamefont {Hoi},
  \citenamefont {Kockum}, \citenamefont {Palomaki}, \citenamefont {Stace},
  \citenamefont {Fan}, \citenamefont {Tornberg}, \citenamefont {Sathyamoorthy},
  \citenamefont {Johansson}, \citenamefont {Delsing},\ and\ \citenamefont
  {Wilson}}]{Hoi2013}%
  \BibitemOpen
  \bibfield  {author} {\bibinfo {author} {\bibfnamefont {I.-C.}\ \bibnamefont
  {Hoi}}, \bibinfo {author} {\bibfnamefont {A.~F.}\ \bibnamefont {Kockum}},
  \bibinfo {author} {\bibfnamefont {T.}~\bibnamefont {Palomaki}}, \bibinfo
  {author} {\bibfnamefont {T.~M.}\ \bibnamefont {Stace}}, \bibinfo {author}
  {\bibfnamefont {B.}~\bibnamefont {Fan}}, \bibinfo {author} {\bibfnamefont
  {L.}~\bibnamefont {Tornberg}}, \bibinfo {author} {\bibfnamefont {S.~R.}\
  \bibnamefont {Sathyamoorthy}}, \bibinfo {author} {\bibfnamefont
  {G.}~\bibnamefont {Johansson}}, \bibinfo {author} {\bibfnamefont
  {P.}~\bibnamefont {Delsing}}, \ and\ \bibinfo {author} {\bibfnamefont
  {C.~M.}\ \bibnamefont {Wilson}},\ }\href {\doibase
  10.1103/PhysRevLett.111.053601} {\bibfield  {journal} {\bibinfo  {journal}
  {Phys.~Rev.~Lett.}\ }\textbf {\bibinfo {volume} {111}},\ \bibinfo {pages}
  {053601} (\bibinfo {year} {2013})}\BibitemShut {NoStop}%
\bibitem [{\citenamefont {van~der Ploeg}\ \emph {et~al.}(2007)\citenamefont
  {van~der Ploeg}, \citenamefont {Izmalkov}, \citenamefont {van~den Brink},
  \citenamefont {H\"ubner}, \citenamefont {Grajcar}, \citenamefont {Il'ichev},
  \citenamefont {Meyer},\ and\ \citenamefont {Zagoskin}}]{vanderPloeg2007}%
  \BibitemOpen
  \bibfield  {author} {\bibinfo {author} {\bibfnamefont {S.~H.~W.}\
  \bibnamefont {van~der Ploeg}}, \bibinfo {author} {\bibfnamefont
  {A.}~\bibnamefont {Izmalkov}}, \bibinfo {author} {\bibfnamefont {A.~M.}\
  \bibnamefont {van~den Brink}}, \bibinfo {author} {\bibfnamefont
  {U.}~\bibnamefont {H\"ubner}}, \bibinfo {author} {\bibfnamefont
  {M.}~\bibnamefont {Grajcar}}, \bibinfo {author} {\bibfnamefont
  {E.}~\bibnamefont {Il'ichev}}, \bibinfo {author} {\bibfnamefont {H.-G.}\
  \bibnamefont {Meyer}}, \ and\ \bibinfo {author} {\bibfnamefont {A.~M.}\
  \bibnamefont {Zagoskin}},\ }\href {\doibase 10.1103/PhysRevLett.98.057004}
  {\bibfield  {journal} {\bibinfo  {journal} {Phys. Rev. Lett.}\ }\textbf
  {\bibinfo {volume} {98}},\ \bibinfo {pages} {057004} (\bibinfo {year}
  {2007})}\BibitemShut {NoStop}%
\bibitem [{\citenamefont {Hime}\ \emph {et~al.}(2006)\citenamefont {Hime},
  \citenamefont {Reichardt}, \citenamefont {Plourde}, \citenamefont
  {Robertson}, \citenamefont {Wu}, \citenamefont {Ustinov},\ and\ \citenamefont
  {Clarke}}]{Hime2006}%
  \BibitemOpen
  \bibfield  {author} {\bibinfo {author} {\bibfnamefont {T.}~\bibnamefont
  {Hime}}, \bibinfo {author} {\bibfnamefont {P.~A.}\ \bibnamefont {Reichardt}},
  \bibinfo {author} {\bibfnamefont {B.~L.~T.}\ \bibnamefont {Plourde}},
  \bibinfo {author} {\bibfnamefont {T.~L.}\ \bibnamefont {Robertson}}, \bibinfo
  {author} {\bibfnamefont {C.-E.}\ \bibnamefont {Wu}}, \bibinfo {author}
  {\bibfnamefont {A.~V.}\ \bibnamefont {Ustinov}}, \ and\ \bibinfo {author}
  {\bibfnamefont {J.}~\bibnamefont {Clarke}},\ }\href {\doibase
  10.1126/science.1134388} {\bibfield  {journal} {\bibinfo  {journal}
  {Science}\ }\textbf {\bibinfo {volume} {314}},\ \bibinfo {pages} {1427}
  (\bibinfo {year} {2006})}\BibitemShut {NoStop}%
\bibitem [{\citenamefont {Yin}\ \emph {et~al.}(2013)\citenamefont {Yin},
  \citenamefont {Chen}, \citenamefont {Sank}, \citenamefont {O'Malley},
  \citenamefont {White}, \citenamefont {Barends}, \citenamefont {Kelly},
  \citenamefont {Lucero}, \citenamefont {Mariantoni}, \citenamefont {Megrant},
  \citenamefont {Neill}, \citenamefont {Vainsencher}, \citenamefont {Wenner},
  \citenamefont {Korotkov}, \citenamefont {Cleland},\ and\ \citenamefont
  {Martinis}}]{Yin:2013}%
  \BibitemOpen
  \bibfield  {author} {\bibinfo {author} {\bibfnamefont {Y.}~\bibnamefont
  {Yin}}, \bibinfo {author} {\bibfnamefont {Y.}~\bibnamefont {Chen}}, \bibinfo
  {author} {\bibfnamefont {D.}~\bibnamefont {Sank}}, \bibinfo {author}
  {\bibfnamefont {P.~J.~J.}\ \bibnamefont {O'Malley}}, \bibinfo {author}
  {\bibfnamefont {T.~C.}\ \bibnamefont {White}}, \bibinfo {author}
  {\bibfnamefont {R.}~\bibnamefont {Barends}}, \bibinfo {author} {\bibfnamefont
  {J.}~\bibnamefont {Kelly}}, \bibinfo {author} {\bibfnamefont
  {E.}~\bibnamefont {Lucero}}, \bibinfo {author} {\bibfnamefont
  {M.}~\bibnamefont {Mariantoni}}, \bibinfo {author} {\bibfnamefont
  {A.}~\bibnamefont {Megrant}}, \bibinfo {author} {\bibfnamefont
  {C.}~\bibnamefont {Neill}}, \bibinfo {author} {\bibfnamefont
  {A.}~\bibnamefont {Vainsencher}}, \bibinfo {author} {\bibfnamefont
  {J.}~\bibnamefont {Wenner}}, \bibinfo {author} {\bibfnamefont {A.~N.}\
  \bibnamefont {Korotkov}}, \bibinfo {author} {\bibfnamefont {A.~N.}\
  \bibnamefont {Cleland}}, \ and\ \bibinfo {author} {\bibfnamefont {J.~M.}\
  \bibnamefont {Martinis}},\ }\href {\doibase 10.1103/PhysRevLett.110.107001}
  {\bibfield  {journal} {\bibinfo  {journal} {Phys.~Rev.~Lett.}\ }\textbf
  {\bibinfo {volume} {110}},\ \bibinfo {pages} {107001} (\bibinfo {year}
  {2013})}\BibitemShut {NoStop}%
\bibitem [{\citenamefont {Pierre}\ \emph {et~al.}(2014)\citenamefont {Pierre},
  \citenamefont {Svensson}, \citenamefont {Raman~Sathyamoorthy}, \citenamefont
  {Johansson},\ and\ \citenamefont {Delsing}}]{Pierre2014}%
  \BibitemOpen
  \bibfield  {author} {\bibinfo {author} {\bibfnamefont {M.}~\bibnamefont
  {Pierre}}, \bibinfo {author} {\bibfnamefont {I.-M.}\ \bibnamefont
  {Svensson}}, \bibinfo {author} {\bibfnamefont {S.}~\bibnamefont
  {Raman~Sathyamoorthy}}, \bibinfo {author} {\bibfnamefont {G.}~\bibnamefont
  {Johansson}}, \ and\ \bibinfo {author} {\bibfnamefont {P.}~\bibnamefont
  {Delsing}},\ }\href {\doibase http://dx.doi.org/10.1063/1.4882646} {\bibfield
   {journal} {\bibinfo  {journal} {Appl. Phys. Lett.}\ }\textbf {\bibinfo
  {volume} {104}},\ \bibinfo {eid} {232604} (\bibinfo {year}
  {2014})}\BibitemShut {NoStop}%
\bibitem [{\citenamefont {Allman}\ \emph {et~al.}(2014)\citenamefont {Allman},
  \citenamefont {Whittaker}, \citenamefont {Castellanos-Beltran}, \citenamefont
  {Cicak}, \citenamefont {da~Silva}, \citenamefont {DeFeo}, \citenamefont
  {Lecocq}, \citenamefont {Sirois}, \citenamefont {Teufel}, \citenamefont
  {Aumentado},\ and\ \citenamefont {Simmonds}}]{Allman2014}%
  \BibitemOpen
  \bibfield  {author} {\bibinfo {author} {\bibfnamefont {M.~S.}\ \bibnamefont
  {Allman}}, \bibinfo {author} {\bibfnamefont {J.~D.}\ \bibnamefont
  {Whittaker}}, \bibinfo {author} {\bibfnamefont {M.}~\bibnamefont
  {Castellanos-Beltran}}, \bibinfo {author} {\bibfnamefont {K.}~\bibnamefont
  {Cicak}}, \bibinfo {author} {\bibfnamefont {F.}~\bibnamefont {da~Silva}},
  \bibinfo {author} {\bibfnamefont {M.~P.}\ \bibnamefont {DeFeo}}, \bibinfo
  {author} {\bibfnamefont {F.}~\bibnamefont {Lecocq}}, \bibinfo {author}
  {\bibfnamefont {A.}~\bibnamefont {Sirois}}, \bibinfo {author} {\bibfnamefont
  {J.~D.}\ \bibnamefont {Teufel}}, \bibinfo {author} {\bibfnamefont
  {J.}~\bibnamefont {Aumentado}}, \ and\ \bibinfo {author} {\bibfnamefont
  {R.~W.}\ \bibnamefont {Simmonds}},\ }\href {\doibase
  10.1103/PhysRevLett.112.123601} {\bibfield  {journal} {\bibinfo  {journal}
  {Phys. Rev. Lett.}\ }\textbf {\bibinfo {volume} {112}},\ \bibinfo {pages}
  {123601} (\bibinfo {year} {2014})}\BibitemShut {NoStop}%
\bibitem [{\citenamefont {Chen}\ \emph {et~al.}(2014)\citenamefont {Chen},
  \citenamefont {Neill}, \citenamefont {Roushan}, \citenamefont {Leung},
  \citenamefont {Fang}, \citenamefont {Barends}, \citenamefont {Kelly},
  \citenamefont {Campbell}, \citenamefont {Chen}, \citenamefont {Chiaro},
  \citenamefont {Dunsworth}, \citenamefont {Jeffrey}, \citenamefont {Megrant},
  \citenamefont {Mutus}, \citenamefont {O'Malley}, \citenamefont {Quintana},
  \citenamefont {Sank}, \citenamefont {Vainsencher}, \citenamefont {Wenner},
  \citenamefont {White}, \citenamefont {Geller}, \citenamefont {Cleland},\ and\
  \citenamefont {Martinis}}]{xmonexp}%
  \BibitemOpen
  \bibfield  {author} {\bibinfo {author} {\bibfnamefont {Y.}~\bibnamefont
  {Chen}}, \bibinfo {author} {\bibfnamefont {C.}~\bibnamefont {Neill}},
  \bibinfo {author} {\bibfnamefont {P.}~\bibnamefont {Roushan}}, \bibinfo
  {author} {\bibfnamefont {N.}~\bibnamefont {Leung}}, \bibinfo {author}
  {\bibfnamefont {M.}~\bibnamefont {Fang}}, \bibinfo {author} {\bibfnamefont
  {R.}~\bibnamefont {Barends}}, \bibinfo {author} {\bibfnamefont
  {J.}~\bibnamefont {Kelly}}, \bibinfo {author} {\bibfnamefont
  {B.}~\bibnamefont {Campbell}}, \bibinfo {author} {\bibfnamefont
  {Z.}~\bibnamefont {Chen}}, \bibinfo {author} {\bibfnamefont {B.}~\bibnamefont
  {Chiaro}}, \bibinfo {author} {\bibfnamefont {A.}~\bibnamefont {Dunsworth}},
  \bibinfo {author} {\bibfnamefont {E.}~\bibnamefont {Jeffrey}}, \bibinfo
  {author} {\bibfnamefont {A.}~\bibnamefont {Megrant}}, \bibinfo {author}
  {\bibfnamefont {J.~Y.}\ \bibnamefont {Mutus}}, \bibinfo {author}
  {\bibfnamefont {P.~J.~J.}\ \bibnamefont {O'Malley}}, \bibinfo {author}
  {\bibfnamefont {C.~M.}\ \bibnamefont {Quintana}}, \bibinfo {author}
  {\bibfnamefont {D.}~\bibnamefont {Sank}}, \bibinfo {author} {\bibfnamefont
  {A.}~\bibnamefont {Vainsencher}}, \bibinfo {author} {\bibfnamefont
  {J.}~\bibnamefont {Wenner}}, \bibinfo {author} {\bibfnamefont {T.~C.}\
  \bibnamefont {White}}, \bibinfo {author} {\bibfnamefont {M.~R.}\ \bibnamefont
  {Geller}}, \bibinfo {author} {\bibfnamefont {A.~N.}\ \bibnamefont {Cleland}},
  \ and\ \bibinfo {author} {\bibfnamefont {J.~M.}\ \bibnamefont {Martinis}},\
  }\href {\doibase 10.1103/PhysRevLett.113.220502} {\bibfield  {journal}
  {\bibinfo  {journal} {Phys. Rev. Lett.}\ }\textbf {\bibinfo {volume} {113}},\
  \bibinfo {pages} {220502} (\bibinfo {year} {2014})}\BibitemShut {NoStop}%
\bibitem [{\citenamefont {Leib}\ \emph {et~al.}(2012)\citenamefont {Leib},
  \citenamefont {Deppe}, \citenamefont {Marx}, \citenamefont {Gross},\ and\
  \citenamefont {Hartmann}}]{Leib2012}%
  \BibitemOpen
  \bibfield  {author} {\bibinfo {author} {\bibfnamefont {M.}~\bibnamefont
  {Leib}}, \bibinfo {author} {\bibfnamefont {F.}~\bibnamefont {Deppe}},
  \bibinfo {author} {\bibfnamefont {A.}~\bibnamefont {Marx}}, \bibinfo {author}
  {\bibfnamefont {R.}~\bibnamefont {Gross}}, \ and\ \bibinfo {author}
  {\bibfnamefont {M.~J.}\ \bibnamefont {Hartmann}},\ }\href
  {http://stacks.iop.org/1367-2630/14/i=7/a=075024} {\bibfield  {journal}
  {\bibinfo  {journal} {New Journal of Physics}\ }\textbf {\bibinfo {volume}
  {14}},\ \bibinfo {pages} {075024} (\bibinfo {year} {2012})}\BibitemShut
  {NoStop}%
\bibitem [{\citenamefont {Hartmann}(2010)}]{Hartmann2010}%
  \BibitemOpen
  \bibfield  {author} {\bibinfo {author} {\bibfnamefont {M.~J.}\ \bibnamefont
  {Hartmann}},\ }\href {\doibase 10.1103/PhysRevLett.104.113601} {\bibfield
  {journal} {\bibinfo  {journal} {Phys. Rev. Lett.}\ }\textbf {\bibinfo
  {volume} {104}},\ \bibinfo {pages} {113601} (\bibinfo {year}
  {2010})}\BibitemShut {NoStop}%
\bibitem [{\citenamefont {Grujic}\ \emph {et~al.}(2013)\citenamefont {Grujic},
  \citenamefont {Clark}, \citenamefont {Jaksch},\ and\ \citenamefont
  {Angelakis}}]{PhysRevA.87.053846}%
  \BibitemOpen
  \bibfield  {author} {\bibinfo {author} {\bibfnamefont {T.}~\bibnamefont
  {Grujic}}, \bibinfo {author} {\bibfnamefont {S.~R.}\ \bibnamefont {Clark}},
  \bibinfo {author} {\bibfnamefont {D.}~\bibnamefont {Jaksch}}, \ and\ \bibinfo
  {author} {\bibfnamefont {D.~G.}\ \bibnamefont {Angelakis}},\ }\href {\doibase
  10.1103/PhysRevA.87.053846} {\bibfield  {journal} {\bibinfo  {journal} {Phys.
  Rev. A}\ }\textbf {\bibinfo {volume} {87}},\ \bibinfo {pages} {053846}
  (\bibinfo {year} {2013})}\BibitemShut {NoStop}%
\bibitem [{\citenamefont {{Gallem{\'{\i}}}}\ \emph {et~al.}(2015)\citenamefont
  {{Gallem{\'{\i}}}}, \citenamefont {{Guilleumas}}, \citenamefont
  {{Martorell}}, \citenamefont {{Mayol}}, \citenamefont {{Polls}},\ and\
  \citenamefont {{Juli{\'a}-D{\'{\i}}az}}}]{Gallemi2015}%
  \BibitemOpen
  \bibfield  {author} {\bibinfo {author} {\bibfnamefont {A.}~\bibnamefont
  {{Gallem{\'{\i}}}}}, \bibinfo {author} {\bibfnamefont {M.}~\bibnamefont
  {{Guilleumas}}}, \bibinfo {author} {\bibfnamefont {J.}~\bibnamefont
  {{Martorell}}}, \bibinfo {author} {\bibfnamefont {R.}~\bibnamefont
  {{Mayol}}}, \bibinfo {author} {\bibfnamefont {A.}~\bibnamefont {{Polls}}}, \
  and\ \bibinfo {author} {\bibfnamefont {B.}~\bibnamefont
  {{Juli{\'a}-D{\'{\i}}az}}},\ }\href
  {http://stacks.iop.org/1367-2630/17/i=7/a=073014} {\bibfield  {journal}
  {\bibinfo  {journal} {New Journal of Physics}\ }\textbf {\bibinfo {volume}
  {17}},\ \bibinfo {pages} {073014} (\bibinfo {year} {2015})}\BibitemShut
  {NoStop}%
\bibitem [{\citenamefont {{Peropadre}}\ \emph {et~al.}(2013)\citenamefont
  {{Peropadre}}, \citenamefont {{Zueco}}, \citenamefont {{Wulschner}},
  \citenamefont {{Deppe}}, \citenamefont {{Marx}}, \citenamefont {{Gross}},\
  and\ \citenamefont {{Garc{\'{\i}}a-Ripoll}}}]{PeropadreTBS}%
  \BibitemOpen
  \bibfield  {author} {\bibinfo {author} {\bibfnamefont {B.}~\bibnamefont
  {{Peropadre}}}, \bibinfo {author} {\bibfnamefont {D.}~\bibnamefont
  {{Zueco}}}, \bibinfo {author} {\bibfnamefont {F.}~\bibnamefont
  {{Wulschner}}}, \bibinfo {author} {\bibfnamefont {F.}~\bibnamefont
  {{Deppe}}}, \bibinfo {author} {\bibfnamefont {A.}~\bibnamefont {{Marx}}},
  \bibinfo {author} {\bibfnamefont {R.}~\bibnamefont {{Gross}}}, \ and\
  \bibinfo {author} {\bibfnamefont {J.~J.}\ \bibnamefont
  {{Garc{\'{\i}}a-Ripoll}}},\ }\href {\doibase 10.1103/PhysRevB.87.134504}
  {\bibfield  {journal} {\bibinfo  {journal} {Phys. Rev. B}\ }\textbf {\bibinfo
  {volume} {87}},\ \bibinfo {eid} {134504} (\bibinfo {year}
  {2013})}\BibitemShut {NoStop}%
\bibitem [{\citenamefont {Tian}\ \emph {et~al.}(2008)\citenamefont {Tian},
  \citenamefont {Allman},\ and\ \citenamefont {Simmonds}}]{ParaCoupQRes}%
  \BibitemOpen
  \bibfield  {author} {\bibinfo {author} {\bibfnamefont {L.}~\bibnamefont
  {Tian}}, \bibinfo {author} {\bibfnamefont {M.~S.}\ \bibnamefont {Allman}}, \
  and\ \bibinfo {author} {\bibfnamefont {R.~W.}\ \bibnamefont {Simmonds}},\
  }\href {\doibase 10.1088/1367-2630/10/11/115001} {\bibfield  {journal}
  {\bibinfo  {journal} {New J. Phys.}\ }\textbf {\bibinfo {volume} {10}},\
  \bibinfo {pages} {115001} (\bibinfo {year} {2008})}\BibitemShut {NoStop}%
\bibitem [{\citenamefont {van~den Brink}\ \emph {et~al.}(2005)\citenamefont
  {van~den Brink}, \citenamefont {Berkley},\ and\ \citenamefont
  {Yalowsky}}]{medTunfluxQcoup}%
  \BibitemOpen
  \bibfield  {author} {\bibinfo {author} {\bibfnamefont {A.~M.}\ \bibnamefont
  {van~den Brink}}, \bibinfo {author} {\bibfnamefont {A.~J.}\ \bibnamefont
  {Berkley}}, \ and\ \bibinfo {author} {\bibfnamefont {M.}~\bibnamefont
  {Yalowsky}},\ }\href {\doibase 10.1088/1367-2630/7/1/230} {\bibfield
  {journal} {\bibinfo  {journal} {New J. Phys.}\ }\textbf {\bibinfo {volume}
  {7}},\ \bibinfo {pages} {230} (\bibinfo {year} {2005})}\BibitemShut {NoStop}%
\bibitem [{\citenamefont {Allman}\ \emph {et~al.}(2010)\citenamefont {Allman},
  \citenamefont {Altomare}, \citenamefont {Whittaker}, \citenamefont {Cicak},
  \citenamefont {Li}, \citenamefont {Sirois}, \citenamefont {Strong},
  \citenamefont {Teufel},\ and\ \citenamefont {Simmonds}}]{RFQResCoupling}%
  \BibitemOpen
  \bibfield  {author} {\bibinfo {author} {\bibfnamefont {M.~S.}\ \bibnamefont
  {Allman}}, \bibinfo {author} {\bibfnamefont {F.}~\bibnamefont {Altomare}},
  \bibinfo {author} {\bibfnamefont {J.~D.}\ \bibnamefont {Whittaker}}, \bibinfo
  {author} {\bibfnamefont {K.}~\bibnamefont {Cicak}}, \bibinfo {author}
  {\bibfnamefont {D.}~\bibnamefont {Li}}, \bibinfo {author} {\bibfnamefont
  {A.}~\bibnamefont {Sirois}}, \bibinfo {author} {\bibfnamefont
  {J.}~\bibnamefont {Strong}}, \bibinfo {author} {\bibfnamefont {J.~D.}\
  \bibnamefont {Teufel}}, \ and\ \bibinfo {author} {\bibfnamefont {R.~W.}\
  \bibnamefont {Simmonds}},\ }\href {\doibase 10.1103/PhysRevLett.104.177004}
  {\bibfield  {journal} {\bibinfo  {journal} {Phys. Rev. Lett.}\ }\textbf
  {\bibinfo {volume} {104}},\ \bibinfo {pages} {177004} (\bibinfo {year}
  {2010})}\BibitemShut {NoStop}%
\bibitem [{\citenamefont {Geller}\ \emph {et~al.}(2015)\citenamefont {Geller},
  \citenamefont {Donate}, \citenamefont {Chen}, \citenamefont {Fang},
  \citenamefont {Leung}, \citenamefont {Neill}, \citenamefont {Roushan},\ and\
  \citenamefont {Martinis}}]{CoupXTh}%
  \BibitemOpen
  \bibfield  {author} {\bibinfo {author} {\bibfnamefont {M.~R.}\ \bibnamefont
  {Geller}}, \bibinfo {author} {\bibfnamefont {E.}~\bibnamefont {Donate}},
  \bibinfo {author} {\bibfnamefont {Y.}~\bibnamefont {Chen}}, \bibinfo {author}
  {\bibfnamefont {M.~T.}\ \bibnamefont {Fang}}, \bibinfo {author}
  {\bibfnamefont {N.}~\bibnamefont {Leung}}, \bibinfo {author} {\bibfnamefont
  {C.}~\bibnamefont {Neill}}, \bibinfo {author} {\bibfnamefont
  {P.}~\bibnamefont {Roushan}}, \ and\ \bibinfo {author} {\bibfnamefont
  {J.~M.}\ \bibnamefont {Martinis}},\ }\href {\doibase
  10.1103/PhysRevA.92.012320} {\bibfield  {journal} {\bibinfo  {journal} {Phys.
  Rev. A}\ }\textbf {\bibinfo {volume} {92}},\ \bibinfo {pages} {012320}
  (\bibinfo {year} {2015})}\BibitemShut {NoStop}%
\bibitem [{\citenamefont {Bourassa}\ \emph {et~al.}(2009)\citenamefont
  {Bourassa}, \citenamefont {Gambetta}, \citenamefont {Abdumalikov},
  \citenamefont {Astafiev}, \citenamefont {Nakamura},\ and\ \citenamefont
  {Blais}}]{USCB}%
  \BibitemOpen
  \bibfield  {author} {\bibinfo {author} {\bibfnamefont {J.}~\bibnamefont
  {Bourassa}}, \bibinfo {author} {\bibfnamefont {J.~M.}\ \bibnamefont
  {Gambetta}}, \bibinfo {author} {\bibfnamefont {A.~A.}\ \bibnamefont
  {Abdumalikov}}, \bibinfo {author} {\bibfnamefont {O.}~\bibnamefont
  {Astafiev}}, \bibinfo {author} {\bibfnamefont {Y.}~\bibnamefont {Nakamura}},
  \ and\ \bibinfo {author} {\bibfnamefont {A.}~\bibnamefont {Blais}},\ }\href
  {\doibase 10.1103/PhysRevA.80.032109} {\bibfield  {journal} {\bibinfo
  {journal} {Phys. Rev. A}\ }\textbf {\bibinfo {volume} {80}},\ \bibinfo
  {pages} {032109} (\bibinfo {year} {2009})}\BibitemShut {NoStop}%
\bibitem [{\citenamefont {{G{\"o}ppl}}\ \emph {et~al.}(2008)\citenamefont
  {{G{\"o}ppl}}, \citenamefont {{Fragner}}, \citenamefont {{Baur}},
  \citenamefont {{Bianchetti}}, \citenamefont {{Filipp}}, \citenamefont
  {{Fink}}, \citenamefont {{Leek}}, \citenamefont {{Puebla}}, \citenamefont
  {{Steffen}},\ and\ \citenamefont {{Wallraff}}}]{Goeppl2008}%
  \BibitemOpen
  \bibfield  {author} {\bibinfo {author} {\bibfnamefont {M.}~\bibnamefont
  {{G{\"o}ppl}}}, \bibinfo {author} {\bibfnamefont {A.}~\bibnamefont
  {{Fragner}}}, \bibinfo {author} {\bibfnamefont {M.}~\bibnamefont {{Baur}}},
  \bibinfo {author} {\bibfnamefont {R.}~\bibnamefont {{Bianchetti}}}, \bibinfo
  {author} {\bibfnamefont {S.}~\bibnamefont {{Filipp}}}, \bibinfo {author}
  {\bibfnamefont {J.~M.}\ \bibnamefont {{Fink}}}, \bibinfo {author}
  {\bibfnamefont {P.~J.}\ \bibnamefont {{Leek}}}, \bibinfo {author}
  {\bibfnamefont {G.}~\bibnamefont {{Puebla}}}, \bibinfo {author}
  {\bibfnamefont {L.}~\bibnamefont {{Steffen}}}, \ and\ \bibinfo {author}
  {\bibfnamefont {A.}~\bibnamefont {{Wallraff}}},\ }\href {\doibase
  10.1063/1.3010859} {\bibfield  {journal} {\bibinfo  {journal} {J. Appl.
  Phys.}\ }\textbf {\bibinfo {volume} {104}},\ \bibinfo {pages} {113904}
  (\bibinfo {year} {2008})}\BibitemShut {NoStop}%
\bibitem [{\citenamefont {Wahl}\ \emph {et~al.}(1999)\citenamefont {Wahl},
  \citenamefont {Schmidt},\ and\ \citenamefont {Forrai}}]{Wahl1999}%
  \BibitemOpen
  \bibfield  {author} {\bibinfo {author} {\bibfnamefont {F.}~\bibnamefont
  {Wahl}}, \bibinfo {author} {\bibfnamefont {G.}~\bibnamefont {Schmidt}}, \
  and\ \bibinfo {author} {\bibfnamefont {L.}~\bibnamefont {Forrai}},\ }\href
  {\doibase 10.1006/jsvi.1998.1831} {\bibfield  {journal} {\bibinfo  {journal}
  {J. Sound Vib.}\ }\textbf {\bibinfo {volume} {219}},\ \bibinfo {pages} {379 } (\bibinfo
  {year} {1999})}\BibitemShut {NoStop}%
\bibitem [{\citenamefont {Sames}\ \emph {et~al.}(2014)\citenamefont {Sames},
  \citenamefont {Chibani}, \citenamefont {Altin}, \citenamefont {Wilk},\ and\
  \citenamefont {Rempe}}]{Antires}%
  \BibitemOpen
  \bibfield  {author} {\bibinfo {author} {\bibfnamefont {C.}~\bibnamefont
  {Sames}}, \bibinfo {author} {\bibfnamefont {H.}~\bibnamefont {Chibani}},
	\bibinfo {author} {\bibfnamefont {C.}\ \bibnamefont {Hamsen}},
  \bibinfo {author} {\bibfnamefont {P.~A.}\ \bibnamefont {Altin}},
  \bibinfo {author} {\bibfnamefont {T.}~\bibnamefont {Wilk}}, \ and\ \bibinfo
  {author} {\bibfnamefont {G.}~\bibnamefont {Rempe}},\ }\href {\doibase
  10.1103/PhysRevLett.112.043601} {\bibfield  {journal} {\bibinfo  {journal}
  {Phys. Rev. Lett.}\ }\textbf {\bibinfo {volume} {112}},\ \bibinfo {pages}
  {043601} (\bibinfo {year} {2014})}\BibitemShut {NoStop}%
\bibitem [{\citenamefont {Zhong}\ \emph {et~al.}(2013)\citenamefont {Zhong},
  \citenamefont {Menzel}, \citenamefont {Candia}, \citenamefont {Eder},
  \citenamefont {Ihmig}, \citenamefont {Baust}, \citenamefont {Haeberlein},
  \citenamefont {Hoffmann}, \citenamefont {Inomata}, \citenamefont {Yamamoto},
  \citenamefont {Nakamura}, \citenamefont {Solano}, \citenamefont {Deppe},
  \citenamefont {Marx},\ and\ \citenamefont {Gross}}]{Zhong}%
  \BibitemOpen
  \bibfield  {author} {\bibinfo {author} {\bibfnamefont {L.}~\bibnamefont
  {Zhong}}, \bibinfo {author} {\bibfnamefont {E.~P.}\ \bibnamefont {Menzel}},
  \bibinfo {author} {\bibfnamefont {R.~D.}\ \bibnamefont {Candia}}, \bibinfo
  {author} {\bibfnamefont {P.}~\bibnamefont {Eder}}, \bibinfo {author}
  {\bibfnamefont {M.}~\bibnamefont {Ihmig}}, \bibinfo {author} {\bibfnamefont
  {A.}~\bibnamefont {Baust}}, \bibinfo {author} {\bibfnamefont
  {M.}~\bibnamefont {Haeberlein}}, \bibinfo {author} {\bibfnamefont
  {E.}~\bibnamefont {Hoffmann}}, \bibinfo {author} {\bibfnamefont
  {K.}~\bibnamefont {Inomata}}, \bibinfo {author} {\bibfnamefont
  {T.}~\bibnamefont {Yamamoto}}, \bibinfo {author} {\bibfnamefont
  {Y.}~\bibnamefont {Nakamura}}, \bibinfo {author} {\bibfnamefont
  {E.}~\bibnamefont {Solano}}, \bibinfo {author} {\bibfnamefont
  {F.}~\bibnamefont {Deppe}}, \bibinfo {author} {\bibfnamefont
  {A.}~\bibnamefont {Marx}}, \ and\ \bibinfo {author} {\bibfnamefont
  {R.}~\bibnamefont {Gross}},\ }\href
  {http://stacks.iop.org/1367-2630/15/i=12/a=125013} {\bibfield  {journal}
  {\bibinfo  {journal} {New Journal of Physics}\ }\textbf {\bibinfo {volume}
  {15}},\ \bibinfo {pages} {125013} (\bibinfo {year} {2013})}\BibitemShut
  {NoStop}%
\bibitem [{\citenamefont {{Menzel}}\ \emph {et~al.}(2012)\citenamefont
  {{Menzel}}, \citenamefont {{Di Candia}}, \citenamefont {{Deppe}},
  \citenamefont {{Eder}}, \citenamefont {{Zhong}}, \citenamefont {{Ihmig}},
  \citenamefont {{Haeberlein}}, \citenamefont {{Baust}}, \citenamefont
  {{Hoffmann}}, \citenamefont {{Ballester}}, \citenamefont {{Inomata}},
  \citenamefont {{Yamamoto}}, \citenamefont {{Nakamura}}, \citenamefont
  {{Solano}}, \citenamefont {{Marx}},\ and\ \citenamefont
  {{Gross}}}]{2012Menzel}%
  \BibitemOpen
  \bibfield  {author} {\bibinfo {author} {\bibfnamefont {E.~P.}\ \bibnamefont
  {{Menzel}}}, \bibinfo {author} {\bibfnamefont {R.}~\bibnamefont {{Di
  Candia}}}, \bibinfo {author} {\bibfnamefont {F.}~\bibnamefont {{Deppe}}},
  \bibinfo {author} {\bibfnamefont {P.}~\bibnamefont {{Eder}}}, \bibinfo
  {author} {\bibfnamefont {L.}~\bibnamefont {{Zhong}}}, \bibinfo {author}
  {\bibfnamefont {M.}~\bibnamefont {{Ihmig}}}, \bibinfo {author} {\bibfnamefont
  {M.}~\bibnamefont {{Haeberlein}}}, \bibinfo {author} {\bibfnamefont
  {A.}~\bibnamefont {{Baust}}}, \bibinfo {author} {\bibfnamefont
  {E.}~\bibnamefont {{Hoffmann}}}, \bibinfo {author} {\bibfnamefont
  {D.}~\bibnamefont {{Ballester}}}, \bibinfo {author} {\bibfnamefont
  {K.}~\bibnamefont {{Inomata}}}, \bibinfo {author} {\bibfnamefont
  {T.}~\bibnamefont {{Yamamoto}}}, \bibinfo {author} {\bibfnamefont
  {Y.}~\bibnamefont {{Nakamura}}}, \bibinfo {author} {\bibfnamefont
  {E.}~\bibnamefont {{Solano}}}, \bibinfo {author} {\bibfnamefont
  {A.}~\bibnamefont {{Marx}}}, \ and\ \bibinfo {author} {\bibfnamefont
  {R.}~\bibnamefont {{Gross}}},\ }\href {\doibase
  10.1103/PhysRevLett.109.250502} {\bibfield  {journal} {\bibinfo  {journal}
  {Physical Review Letters}\ }\textbf {\bibinfo {volume} {109}},\ \bibinfo
  {pages} {250502} (\bibinfo {year} {2012})}\BibitemShut {NoStop}%
\bibitem [{\citenamefont {Yamamoto}\ \emph {et~al.}(2008)\citenamefont
  {Yamamoto}, \citenamefont {Inomata}, \citenamefont {Watanabe}, \citenamefont
  {Matsuba}, \citenamefont {Miyazaki}, \citenamefont {Oliver}, \citenamefont
  {Nakamura},\ and\ \citenamefont {Tsai}}]{JPA}%
  \BibitemOpen
  \bibfield  {author} {\bibinfo {author} {\bibfnamefont {T.}~\bibnamefont
  {Yamamoto}}, \bibinfo {author} {\bibfnamefont {K.}~\bibnamefont {Inomata}},
  \bibinfo {author} {\bibfnamefont {M.}~\bibnamefont {Watanabe}}, \bibinfo
  {author} {\bibfnamefont {K.}~\bibnamefont {Matsuba}}, \bibinfo {author}
  {\bibfnamefont {T.}~\bibnamefont {Miyazaki}}, \bibinfo {author}
  {\bibfnamefont {W.~D.}\ \bibnamefont {Oliver}}, \bibinfo {author}
  {\bibfnamefont {Y.}~\bibnamefont {Nakamura}}, \ and\ \bibinfo {author}
  {\bibfnamefont {J.~S.}\ \bibnamefont {Tsai}},\ }\href {\doibase
  10.1063/1.2964182} {\bibfield  {journal} {\bibinfo  {journal} {Appl. Phys.
  Lett.}\ }\textbf {\bibinfo {volume} {93}},\ \bibinfo {pages} {042510}
  (\bibinfo {year} {2008})}\BibitemShut {NoStop}%
\bibitem [{\citenamefont {Bergeal}\ \emph {et~al.}(2010)\citenamefont
  {Bergeal}, \citenamefont {Schackert}, \citenamefont {Metcalfe}, \citenamefont
  {Vijay}, \citenamefont {Manucharyan}, \citenamefont {Frunzio}, \citenamefont
  {Prober}, \citenamefont {Schoelkopf}, \citenamefont {Girvin},\ and\
  \citenamefont {Devoret}}]{JPC}%
  \BibitemOpen
  \bibfield  {author} {\bibinfo {author} {\bibfnamefont {N.}~\bibnamefont
  {Bergeal}}, \bibinfo {author} {\bibfnamefont {F.}~\bibnamefont {Schackert}},
  \bibinfo {author} {\bibfnamefont {M.}~\bibnamefont {Metcalfe}}, \bibinfo
  {author} {\bibfnamefont {R.}~\bibnamefont {Vijay}}, \bibinfo {author}
  {\bibfnamefont {V.~E.}\ \bibnamefont {Manucharyan}}, \bibinfo {author}
  {\bibfnamefont {L.}~\bibnamefont {Frunzio}}, \bibinfo {author} {\bibfnamefont
  {D.~E.}\ \bibnamefont {Prober}}, \bibinfo {author} {\bibfnamefont {R.~J.}\
  \bibnamefont {Schoelkopf}}, \bibinfo {author} {\bibfnamefont {S.~M.}\
  \bibnamefont {Girvin}}, \ and\ \bibinfo {author} {\bibfnamefont {M.~H.}\
  \bibnamefont {Devoret}},\ }\href {\doibase 10.1038/nature09035} {\bibfield
  {journal} {\bibinfo  {journal} {Nature}\ }\textbf {\bibinfo {volume} {465}},\
  \bibinfo {pages} {64} (\bibinfo {year} {2010})}\BibitemShut {NoStop}%
\bibitem [{\citenamefont {{Yurke}}\ and\ \citenamefont
  {{Buks}}(2006)}]{2006JPATheo}%
  \BibitemOpen
  \bibfield  {author} {\bibinfo {author} {\bibfnamefont {B.}~\bibnamefont
  {{Yurke}}}\ and\ \bibinfo {author} {\bibfnamefont {E.}~\bibnamefont
  {{Buks}}},\ }\href {\doibase 10.1109/JLT.2006.884490} {\bibfield  {journal}
  {\bibinfo  {journal} {J.Lightw. Technol.}\ }\textbf {\bibinfo {volume}
  {24}},\ \bibinfo {pages} {5054} (\bibinfo {year} {2006})}\BibitemShut
  {NoStop}%
\bibitem [{\citenamefont {Castellanos-Beltran}\ \emph
  {et~al.}(2009)\citenamefont {Castellanos-Beltran}, \citenamefont {Irwin},
  \citenamefont {Vale}, \citenamefont {Hilton},\ and\ \citenamefont
  {Lehnert}}]{arrayJPA}%
  \BibitemOpen
  \bibfield  {author} {\bibinfo {author} {\bibfnamefont {M.}~\bibnamefont
  {Castellanos-Beltran}}, \bibinfo {author} {\bibfnamefont {K.}~\bibnamefont
  {Irwin}}, \bibinfo {author} {\bibfnamefont {L.}~\bibnamefont {Vale}},
  \bibinfo {author} {\bibfnamefont {G.}~\bibnamefont {Hilton}}, \ and\ \bibinfo
  {author} {\bibfnamefont {K.}~\bibnamefont {Lehnert}},\ }\href {\doibase
  10.1109/TASC.2009.2018119} {\bibfield  {journal} {\bibinfo  {journal} {IEEE
  Trans. Appl. Supercond.}\ }\textbf {\bibinfo {volume} {19}},\ \bibinfo
  {pages} {944} (\bibinfo {year} {2009})}\BibitemShut {NoStop}%
\bibitem [{\citenamefont {Houck}\ \emph {et~al.}(2012)\citenamefont {Houck},
  \citenamefont {Tureci},\ and\ \citenamefont {Koch}}]{Houck2012}%
  \BibitemOpen
  \bibfield  {author} {\bibinfo {author} {\bibfnamefont {A.~A.}\ \bibnamefont
  {Houck}}, \bibinfo {author} {\bibfnamefont {H.~E.}\ \bibnamefont {Tureci}}, \
  and\ \bibinfo {author} {\bibfnamefont {J.}~\bibnamefont {Koch}},\ }\href
  {http://dx.doi.org/10.1038/nphys2251} {\bibfield  {journal} {\bibinfo
  {journal} {Nat Phys}\ }\textbf {\bibinfo {volume} {8}},\ \bibinfo {pages}
  {292} (\bibinfo {year} {2012})}\BibitemShut {NoStop}%
\bibitem [{\citenamefont {Raftery}\ \emph {et~al.}(2014)\citenamefont
  {Raftery}, \citenamefont {Sadri}, \citenamefont {Schmidt}, \citenamefont
  {T\"ureci},\ and\ \citenamefont {Houck}}]{Raftery2013}%
  \BibitemOpen
  \bibfield  {author} {\bibinfo {author} {\bibfnamefont {J.}~\bibnamefont
  {Raftery}}, \bibinfo {author} {\bibfnamefont {D.}~\bibnamefont {Sadri}},
  \bibinfo {author} {\bibfnamefont {S.}~\bibnamefont {Schmidt}}, \bibinfo
  {author} {\bibfnamefont {H.~E.}\ \bibnamefont {T\"ureci}}, \ and\ \bibinfo
  {author} {\bibfnamefont {A.~A.}\ \bibnamefont {Houck}},\ }\href {\doibase
  10.1103/PhysRevX.4.031043} {\bibfield  {journal} {\bibinfo  {journal} {Phys.
  Rev. X}\ }\textbf {\bibinfo {volume} {4}},\ \bibinfo {pages} {031043}
  (\bibinfo {year} {2014})}\BibitemShut {NoStop}%

\end{thebibliography}
\end{document}